\documentclass[12pt,twocolumn,tighten]{aastex63}
\usepackage{amsmath,amstext,amssymb}
\usepackage[T1]{fontenc}
\usepackage{apjfonts}
\usepackage[figure,figure*]{hypcap}
\usepackage{graphics,graphicx}
\usepackage{hyperref}
\usepackage{natbib}
\usepackage[caption=false]{subfig} 
\usepackage{enumitem} 
\usepackage{epigraph}

\newcommand{\cn}{NGC\,2516} 

\newcommand{\nkinematic}{3{,}298} 
\newcommand{\nnbhd}{13{,}843} 
\newcommand{\ncore}{1{,}106}  
\newcommand{\nhalo}{2{,}192} 

\newcommand{\ncdips}{2{,}205}
\newcommand{\nkinematicrpltsixteen}{2{,}270}

\newcommand{\numsouthernuniqlcs}{483{,}407} 

\newcommand{\ncalibration}{9{,}619{,}784} 
\newcommand{\nnbhdcalibstar}{1{,}987} 

\newcommand{\nautorotdenominator}{1{,}641} 
\newcommand{\nautorotnumerator}{987} 
\newcommand{\nautorotnumeratormatching}{892} 

\newcommand{\ncompstardenominator}{656} 
\newcommand{\ncompstarnumerator}{107} 
\newcommand{\ncompfrac}{16.3\%} 

\newcommand{\nautovscompstardenominator}{327} 
\newcommand{\nautovscompstarnumerator}{164} 
\newcommand{\nautofrac}{50.1\%} 

\newcommand{\kms}{\,km\,s$^{-1}$}

\newcommand{\bpmrpo}{(G_{\rm BP}-G_{\rm RP})_0}
\newcommand{\bpmrp}{G_{\rm BP}-G_{\rm RP}}

\received{April 29, 2021}
\revised{July 7, 2021}
\accepted{July 16, 2021}
\submitjournal{AAS Journals}
\shorttitle{The Halo of NGC\,2516}

\begin{document}

\defcitealias{bouma_wasp4b_2019}{B19}
\defcitealias{cantatgaudin_gaia_2018}{CG18}
\defcitealias{kounkel_untangling_2019}{KC19}
\defcitealias{meingast_2021}{M21}

\title{
  Rotation and Lithium Confirmation of a 500 Parsec Halo for the Open
  Cluster NGC\,2516\footnote{This is paper 3 of the Cluster Difference
  Imaging Photometric Survey.}
}

\correspondingauthor{L.\,G.\,Bouma}
\email{luke@astro.princeton.edu}

\author[0000-0002-0514-5538]{L. G. Bouma}
\affiliation{Department of Astrophysical Sciences, Princeton University, 4 Ivy Lane, Princeton, NJ 08540, USA}

\author[0000-0002-2792-134X]{J. L. Curtis}
\affiliation{Department of Astronomy, Columbia University, 550 West 120th Street, New York, NY 10027, USA}
\affiliation{Department of Astrophysics, American Museum of Natural History, Central Park West, New York, NY 10024, USA}

\author[0000-0001-8732-6166]{J. D. Hartman}
\affiliation{Department of Astrophysical Sciences, Princeton University, 4 Ivy Lane, Princeton, NJ 08540, USA}

\author[0000-0002-4265-047X]{J. N. Winn}
\affiliation{Department of Astrophysical Sciences, Princeton University, 4 Ivy Lane, Princeton, NJ 08540, USA}

\author[0000-0001-7204-6727]{G. \'A. Bakos}
\affiliation{Department of Astrophysical Sciences, Princeton University, 4 Ivy Lane, Princeton, NJ 08540, USA}
\affiliation{Institute for Advanced Study, 1 Einstein Drive, Princeton, NJ 08540, USA}

%
%
\begin{abstract}
Recent analyses of the Gaia data have identified diffuse stellar
populations surrounding nearby open clusters.  It is important to
verify that these ``halos'', ``tails'', and ``strings'' are of
similar ages and compositions as stars in the denser part of the
cluster.  We present an analysis of NGC\,2516 ($\approx$150\,Myr),
which has a classical tidal radius of 10\,pc and an apparent halo of
stars spanning 500\,pc ($20^\circ$ on-sky).  Combining photometry
from Gaia, rotation periods from TESS, and lithium measurements from
Gaia-ESO and GALAH, we find that the halo of NGC\,2516 is the same
age as the cluster's core.  Two thirds of kinematically selected
halo members out to 250\,pc from the cluster center have rotation
periods consistent with a gyrochronological age of 150\,Myr.  A
comparison sample of field stars shows no such trend.  The lithium
abundances of stars in the halo are higher than in the field, and
are correlated with the stellar rotation rate and binarity fraction,
as has been noted in other young open clusters.  Broadly speaking,
this work supports a new paradigm wherein the halos of open clusters
are often more populous than their cores.  We highlight implications
for spectroscopic survey targeting, open cluster dispersal, and
planet searches around young stars.
\end{abstract}



\section{Introduction}

Star clusters form when dense filaments in hierarchically structured
molecular clouds collide and collapse \citep{shu_star_1987}.  Over the
first 10\,Myr, feedback effects including protostellar outflows,
photoionization, radiation pressure, and supernova shocks disperse the
gas \citep{krumholz_star_2019}.  Since the gas represents about 99\%
of the mass of the original cloud, gas dispersal enables the majority
($\sim$90\%) of stars in the cluster to escape the cluster's
gravitational well \citep{lada_embedded_2003}.

From 10 to 1000\,Myr, the cluster remnants that survive gas dispersal
suffer an onslaught of supernovae, AGB winds, and close stellar
encounters that almost always leads to dissolution
\citep{lamers_mass_loss_2010}.  Extrinsic to the cluster, collisions
with giant molecular clouds \citep{spitzer_disruption_1958}, and
perturbations from the galactic tide in both the radial and vertical
dimensions further promote stellar escape \citep[{\it
e.g.},][]{fukushige_timescale_2000,bergond_gravitational_2001}.

Finding the stars that have dispersed from their clusters into the
galactic field is a pressing topic for a few reasons.  One is to
understand the spatial extent and duration of star formation events
\citep[{\it e.g.},][]{wright_kinematics_2018}.  Another is to
understand the Sun's birth environment \citep{adams_birth_2010}.  The
plausibility of finding the stars that formed with the Sun, whether
through chemical tagging
\citep{freeman_new_2002,hogg_chemical_2016,ness_dopplergangers_2018},
kinematic analyses, or both, can be tested by first using the
remnants of open clusters that have only recently
dissipated into the field.

A separate project that benefits from the new discoveries of dispersed
stellar populations is that of finding young transiting exoplanets
\citep[{\it
e.g.},][]{Mann_K2_33b_2016,ciardi_k2-136_2018,david_four_2019,livingston_k2-264_2019,bouma_cluster_2020,rizzuto_tess_2020,plavchan_planet_2020,newton_2021,tofflemire_2021,zhou_2021_tois}.
Young transiting planets are hard to find because young stars are
rare, and mostly reside in the crowded galactic plane
\citep{Kharchenko_et_al_2013,piskunov_global_2018}.  If the dispersed
halos of nearby star clusters could be reliably identified, this could
expand the census of nearby young stars by up to a factor of $10$,
based on the expected fraction of stars thought to be lost during gas
dispersal.

Although it has been possible to detect dispersed stellar associations
for a long time, Gaia is now enabling major advances \citep[{\it
e.g.},][]{de_zeeuw_hipparcos_1999,bergond_gravitational_2001,zuckerman_young_2004,oh_comoving_2017,cantatgaudin_gaia_2018,gagne_banyan_XII_2018,gagne_banyan_XIII_2018,kounkel_apogee2_2018,zari_3d_2018,kounkel_untangling_2019,furnkranz_2019}.
The populations found by the most recent studies can be summarized as
follows.  On one end are groups with no discernable cores ({\it
e.g.}, Psc-Eri and $\mu$-Tau,
\citealt{meingast_psceri_2019,curtis_tess_2019,gagne_mutau_2020}).  On
the other are hierarchically structured associations with many over
and under-densities  \citep[{\it e.g.} the Sco-Cen and Vela
associations,][]{pecaut_star_2016,cantatgaudin_velaOB2_2019}.  Here,
we focus on an intermediate regime: low-density halos associated with
a single overdensity, typically an open cluster that was known before
Gaia
\citep[see][]{kounkel_untangling_2019,kounkel_untanglingII_2020,meingast_2021}.
In some cases, these halos may actually be tidal tails, as have been
reported in the Hyades \citep{meingast_hyades_2019,roser_hyades_2019},
the Ursa Major moving group \citep{gagne_lowmassUMA_2020}, and Coma
Berenices \citep{tang_comaber_2019,furnkranz_2019}. 

One point of difficulty however is that different algorithms for
identifying clusters yield differing levels of sensitivity and
precision \citep{hunt_clustering_2020}.  For example, using a Gaussian
Mixture Model tautologically yields open clusters that are elliptical
\citep[{\it e.g.},][]{wallace_m4_2018}.  Some unsupervised methods
yield dispersed and asymmetric structures with number densities down
to 100 times lower than the field around the same regions ({\it
e.g.}, \citealt{kounkel_untangling_2019} and \citealt{meingast_2021}).
Since different techniques yield quantitatively and
qualitatively different outcomes, the purity of kinematically selected
samples is important to verify through independent lines of evidence,
including rotation periods and spectroscopy.

As part of a Cluster Difference Imaging Photometric Survey (CDIPS,
\citealt{bouma_cdipsI_2019}), we have been making TESS light
curves of stars reported to be members of coeval populations and
searching them for planets \citep{bouma_cluster_2020}.  Our analysis
of TESS Sectors 1-13 yielded light curves of \numsouthernuniqlcs\
candidate cluster members in the Southern Ecliptic hemisphere, which
are available on
MAST.\footnote{\url{https://archive.stsci.edu/hlsp/cdips}}

As part of this project, we focus here on a single southern open
cluster, and ask: is the cluster halo truly coeval with the core?  We
chose \cn\ for this analysis because it is young (100-200\,Myr) and
close enough ($d\approx400$\,pc) to facilitate measurements of stellar rotation periods using TESS.  \citet{healy_stellar_2020}
in fact already used TESS rotation periods to study the stellar inclination
distribution of stars in the core of NGC\,2516.
The cluster's halo however was reported by
\citet{kounkel_untangling_2019} to span
$\approx20\times10\times350$\,pc, which is a larger volume than
has been considered in previous rotation analyses
\citep[][]{Irwin_NGC2516_2007,fritzewski_rotation_2020,healy_stellar_2020}.
We want to know: is this halo real or is it just a collection of
interlopers, foreground and background stars?  To what extent can we
use Gaia alone to reliably identify age-dated needles in the haystack
of field stars?  What are the implications, observationally and
theoretically, if we can identify the dispersed halos of open
clusters?

To assess whether the stars in the cluster's halo are the same
ages as stars in the cluster's core, we apply three different age-dating
techniques: isochrones, gyrochrones, and lithium depletion
\citep[see][for reviews]{soderblom_ages_2010,soderblom_ages_2014}.
Gyrochronology has now been empirically calibrated for FGK stars with
ages spanning $\approx$10\,Myr to $\approx$4\,Gyr \citep[{\it
e.g.},][]{rebull_usco_2018,curtis_temporary_2019,curtis_rup147_2020,barnes_rotation_2016}.
Lithium equivalent widths can be used for relative age-dating of dwarf
stars over similar age ranges, though for a narrower range of spectral
types \citep{sestito_2005}.  One complication when deriving relative
lithium ages of the K dwarfs comes from the correlation between
lithium abundance and stellar rotation rate, which has recently been
reviewed by \citet{bouvier_lithium-rotation_2020}, and is a subject
that we explore in the latter portions of our analysis.  

A brief note on terminology.  Low-density stellar associations
connected to a dense population (a ``core'') have been described as
being in ``halos'', ``strings'', ``coronae'', ``snakes'',
``outskirts'', and ``tidal tails'' \citep[{\it
e.g.},][]{chumak_tails_2006,davenport_death_2010,kounkel_untangling_2019,roser_hyades_2019,tian_discovery_2020,meingast_2021}.
The latter term implies a particular model for the formation of the
dispersed group.  ``Halo'' is model agnostic, but is not
an ideal term because it can connote spherical symmetry, which is rarely
the case.  The halo of \cn\ is most accurately described as consisting
of a leading and trailing tail.  We call it a halo for simplicity.

Section~\ref{sec:gaia} summarizes the astrometric analyses of the Gaia
data that led to our interest in \cn.  Section~\ref{sec:agedate}
measures the age of the cluster's halo and core using Gaia photometry
(Section~\ref{subsec:hr}), TESS gyrochronology
(Section~\ref{subsec:tess}), and lithium equivalent widths
(Section~\ref{subsec:lithium}).  In Section~\ref{sec:discussion} we
discuss the implications of our analysis for open cluster dispersal,
stellar spin-down, and the lithium-rotation correlation.  Section~\ref{sec:conclusion}
gives our conclusions.

\section{A Dispersed Halo and a Core?}
\label{sec:gaia}

\begin{figure*}[t]
	\begin{center}
		\leavevmode
		\includegraphics[width=0.9\textwidth]{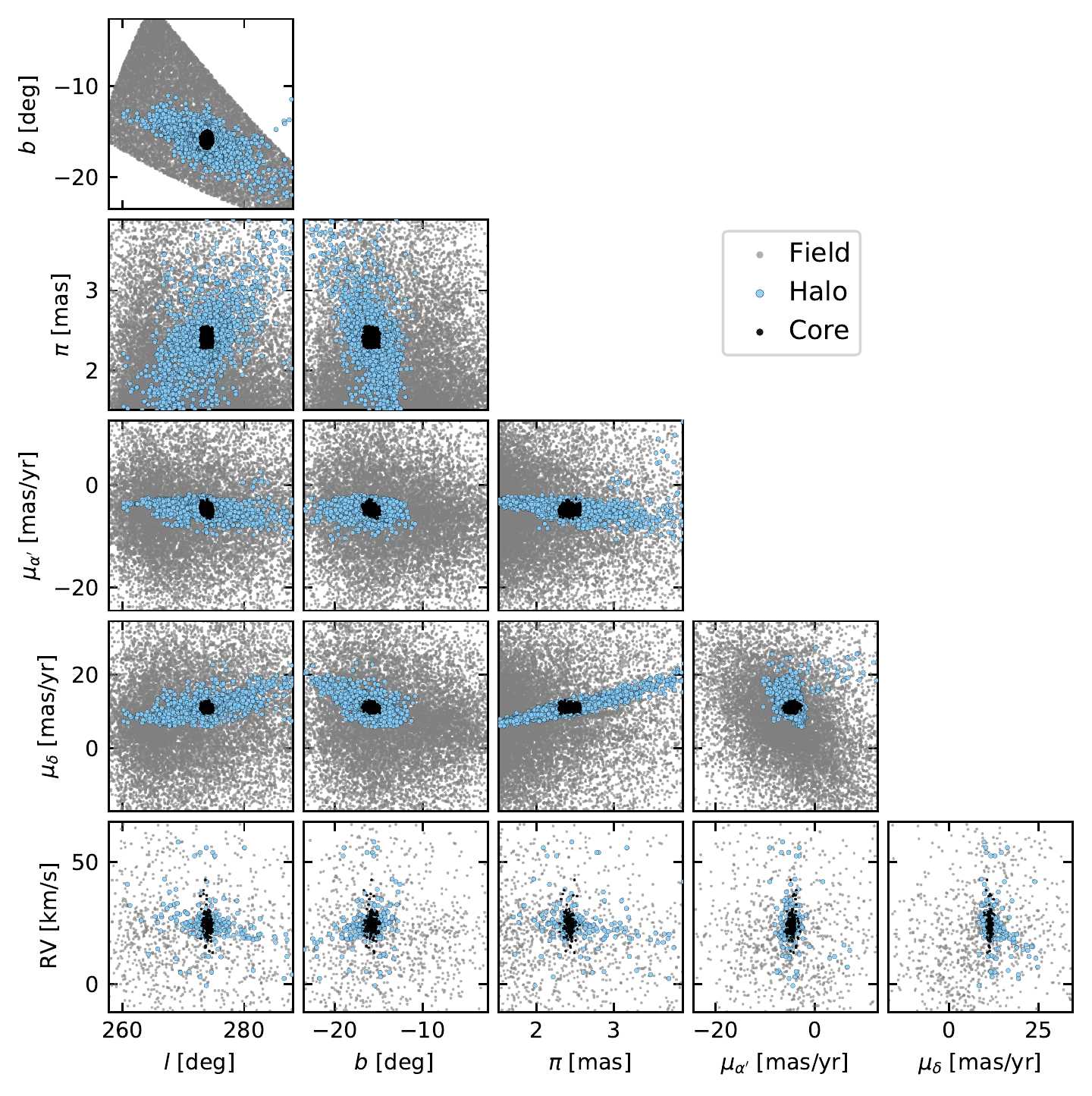}
	\end{center}
	\vspace{-0.7cm}
  \caption{ {\bf The dense and diffuse components of NGC\,2516 across
  position and velocity space.} The core (black) was analyzed by
  \citet{cantatgaudin_gaia_2018} using Gaia DR2, and is coincident
  with the visual overdensity of stars canonically accepted as ``the
  cluster''.  The halo (blue) is a concatenation of studies by
  \citet{kounkel_untangling_2019} and \citet{meingast_2021}, which
  used less restrictive membership assignment algorithms described in
  Appendix~\ref{app:clustering}.  The field sample is randomly drawn
  from an $(\alpha, \delta, \pi)$ cube centered on the cluster.  The
  low volume density of the halo stars makes it difficult to visually
  distinguish the field and the halo samples.
  \label{fig:gaia6d}
	}
\end{figure*}

\begin{figure*}[t]
	\begin{center}
		\leavevmode
		\includegraphics[width=\textwidth]{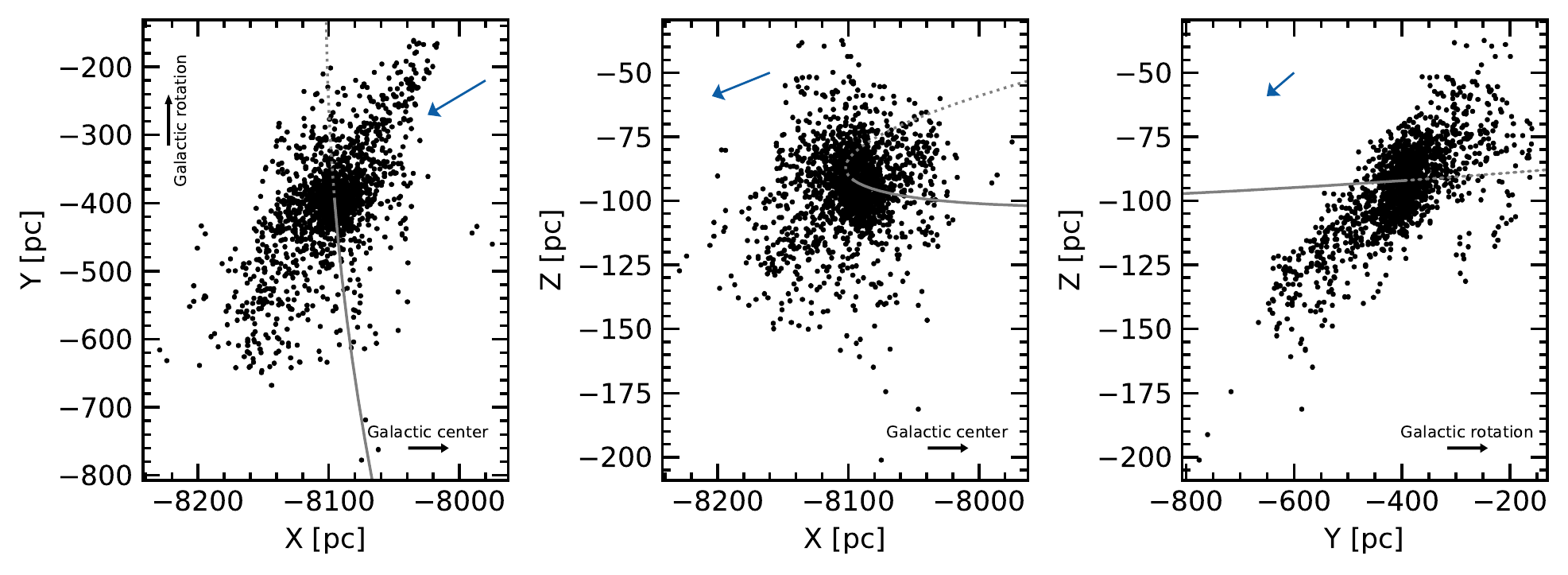}
	\end{center}
	\vspace{-0.7cm}
	\caption{ {\bf Position and orbit of NGC\,2516 in the Galaxy.}
		Points are halo members with $\pi/\sigma_\pi>20$; the cluster is
		most elongated viewed top-down ({\it left}). 
		The Sun is at $\{X,Y,Z\} = \{ -8122, 0, +20.8\}\,{\rm pc}$.
		Gray lines trace the past (solid) and future (dashed)
		cluster orbit, and blue arrows denote the mean cluster
		velocity after subtracting the local standard of rest: $\{v_{\rm X},
		v_{\rm Y}, v_{\rm Z}\} = \{-22.2, -25.3, -4.6\}$\kms. 
		\label{fig:XYZ}
	}
\end{figure*}

We selected candidate \cn\ members based on those reported in the
literature.  While some pre-Gaia analyses were available
\citep{jeffries_ngc2516_2001,Kharchenko_et_al_2013}, the purity and
accuracy of the Gaia-derived results are the current state of the art.
We therefore adopted what we viewed as the most important Gaia-based
samples to compare: those of \citet{cantatgaudin_gaia_2018},
\citet{kounkel_untangling_2019} and \citet{meingast_2021}.  We refer
to these studies in the following as
\citetalias{cantatgaudin_gaia_2018},
\citetalias{kounkel_untangling_2019}, and \citetalias{meingast_2021}
respectively.  A useful visualization of these samples is
available online.\footnote{
  \url{https://homepage.univie.ac.at/stefan.meingast/coronae.html},
  made by \citet{meingast_2021}, last accessed \texttt{2021/07/01}.}

The Gaia clustering studies each used different selection functions,
which yielded different results.  \citetalias{cantatgaudin_gaia_2018}
considered stars brighter than $G=18$\,mag within a few degrees of the
center of \cn, and reported 1106 candidate cluster members.
\citetalias{kounkel_untangling_2019} and \citetalias{meingast_2021}
considered stars up to $\approx$1\,mag fainter, and reported 3003 and
1860 members respectively.  The unsupervised clustering techniques
that each of these studies applied to the second Gaia data release are
discussed in Appendix~\ref{app:clustering}, as is the overlap between
their resulting membership catalogs.

Figure~\ref{fig:gaia6d} shows the cluster members reported by each
study in the space of observed positions, proper motions, and radial
velocity when available.
In Figure~\ref{fig:gaia6d}, and for the remainder of the study, we
describe the \citetalias{cantatgaudin_gaia_2018} members as the
``core'' of the cluster, and any non-overlapping
\citetalias{kounkel_untangling_2019} and \citetalias{meingast_2021}
members as the ``halo''.  This defintion implies that there are
\ncore\ core members, and \nhalo\ halo members.  The distinction is
tautological, in the sense that \citetalias{cantatgaudin_gaia_2018} did
not extend their search for members out to tens of degrees from the
cluster center.  Nonetheless, the \citetalias{cantatgaudin_gaia_2018}
membership catalog is consistent with that of many earlier
investigators \citep[{\it
e.g.},][]{jeffries_ngc2516_2001,Kharchenko_et_al_2013}, and is
consistent with the general visual overdensity one sees when viewing
\cn\ in the sky: it appears to span $\approx$2$^\circ$, not
$\approx20^\circ$.

Outside of the core and halo, we also define a set of nearby field
stars in the ``neighborhood'' of \cn.  Based on the observed
distribution of halo members, we drew these stars randomly from the
following intervals of right ascension, declination, and parallax:
\begin{align}
  \alpha\,[\mathrm{deg}] &\in [108, 132], \\
  \delta\,[\mathrm{deg}] &\in [-76, -45], \\
  \pi\,[\mathrm{mas}] &\in [1.5, 4.0].
\end{align}
We imposed a magnitude limit of $G=19$\,mag, and ran the queries using
the \texttt{astroquery.gaia} module \citep{astroquery_2018}.  We
allowed the number of stars in the comparison sample to exceed that in
the cluster sample by a factor of $\approx$5, to ensure broad sampling
of stellar masses and evolutionary states.  We also required the
comparison sample to not overlap with the cluster sample.

The style of visualization given in Figure~\ref{fig:gaia6d} does not
make it clear, in our eyes, why the clustering algorithms have decided
to associate certain stars with the halo.  Canonically, open clusters
are spherical, and span $\approx$10\,pc \citep[{\it
e.g.},][]{Kharchenko_et_al_2013}.  Inverting the parallaxes in
Figure~\ref{fig:gaia6d} shows that members have been reported from
line-of-sight distances ranging from $\approx$$300$ to
$\approx$$600\,$pc.  Is this structure real? What explains its
existence?

An initial step in visualizing the structure and kinematics
of the candidate cluster members is to consider their Cartesian
coordinates (Figure~\ref{fig:XYZ}).  We computed these positions by
transforming from $(\alpha, \delta, \pi)$ to galactic ($X,Y,Z$)
assuming the \texttt{astropy v4.0} coordinate standard
\citep{astropy_2018}.  The Galaxy rotates in the
$+\hat{Y}$ direction. The cluster orbit (gray line) was evaluated by taking the
median parameters for core members for which
\citetalias{cantatgaudin_gaia_2018} reported membership probabilities
exceeding 70\%.  We then integrated the orbit using \texttt{gala} and
the \texttt{MWPotential2014} potential \citep{bovy_galpy_2015,gala}.

The general shape of the cluster itself seems to include a central
overdensity and two tails (the halo).  One tail is leading the
cluster's orbit, and is angled toward the center of the galaxy when
viewed top-down.  The other tail is trailing the cluster's orbit, and
is pointed away from the center of the galaxy.  The elongation of the
cluster in both the $(X,Y)$ and $(Z,Y)$ planes is correlated with the
direction of the LSR-corrected median cluster velocity.   Possible
explanations for this overall morphology are discussed in
Section~\ref{subsec:origin}.

\section{Age-Dating the Halo of NGC\,2516}
\label{sec:agedate}

\subsection{HR Diagram from Gaia}
\label{subsec:hr}

\begin{figure*}[t]
	\begin{center}
		\leavevmode
		\subfloat{
			\includegraphics[width=0.49\textwidth]{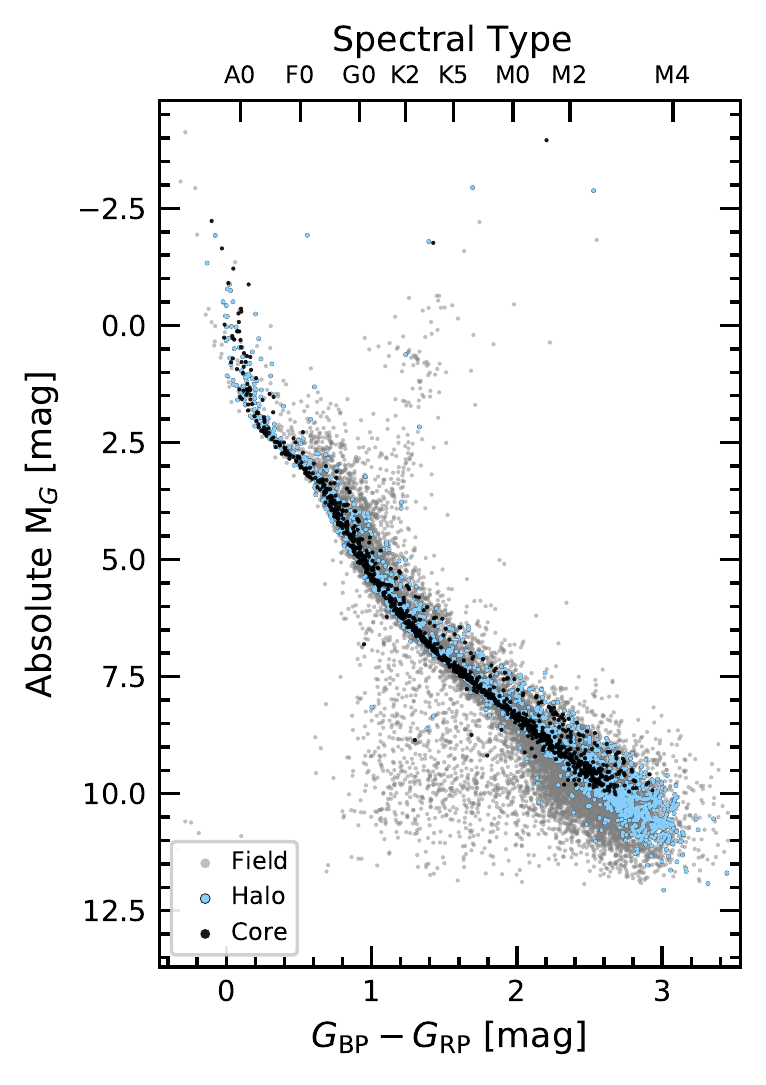}
			\includegraphics[width=0.49\textwidth]{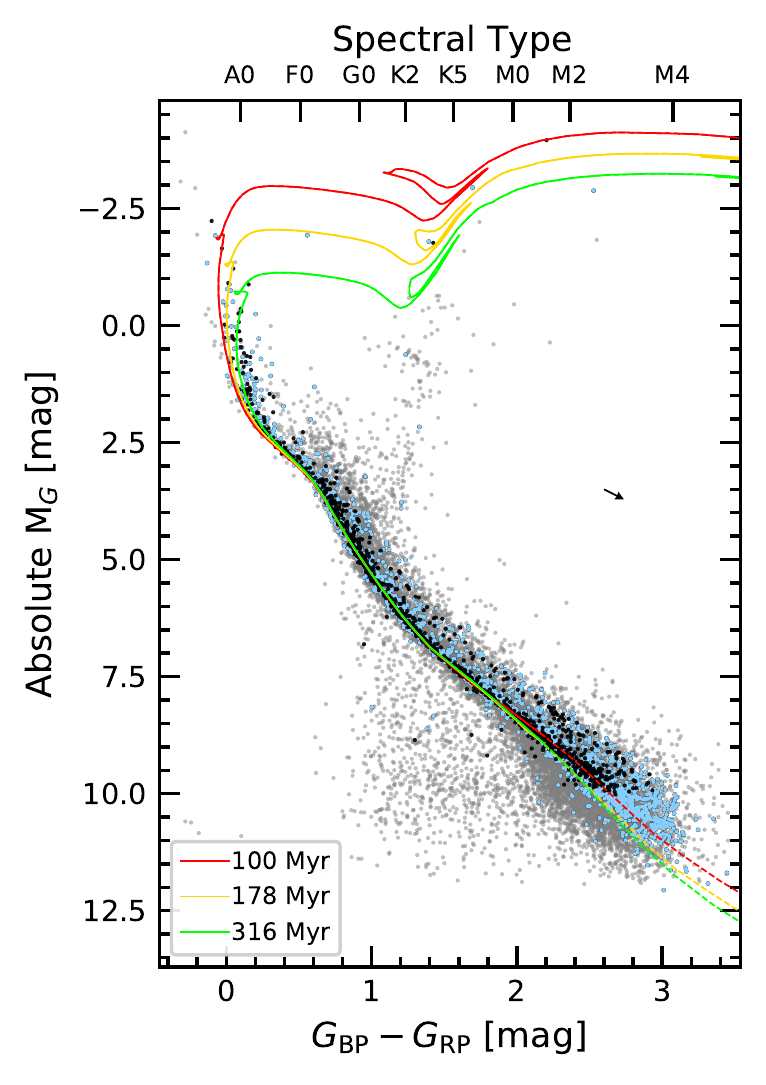}
		}
	\end{center}
	\vspace{-0.7cm}
  \caption{ {\bf HR diagrams of NGC\,2516.}
    {\it Left}: Gaia EDR3 photometry for the core (black) shows a
    main-sequence and turnoff consistent with stars of a fixed age and
    metallicity.  The halo (blue) is similar, though the binary
    sequence is less defined.  The faintest M dwarfs in the core and
    halo are brighter than field stars (gray) with the same color,
    consistent with the cluster M dwarfs not having yet reached the
    main-sequence.  {\it Right}: PARSEC isochrones ([Fe/H]=+0.06) are
    overplotted at intervals of $\log_{10}(t/\mathrm{yr})=\{8, 8.25,
    8.5\}$.  The arrow represents the average reddening correction
    applied to the models.  The opacity of the M-dwarf model
    atmospheres is less well-calibrated for $M_\star \lesssim
    0.45M_\odot$ ($\approx$M2V); these model values are shown with
    dashed lines.
    \label{fig:hr}
  }
\end{figure*}

The first check on whether the candidate cluster members share the
same age is to analyze the Gaia Hertzsprung-Russell (HR) diagrams.
Comparable analyses have previously been performed by
\citetalias{cantatgaudin_gaia_2018},
\citetalias{kounkel_untangling_2019}, and \citetalias{meingast_2021}.

Figure~\ref{fig:hr} shows the HR diagram in the space of absolute Gaia
$G$ magnitude as a function of observed $\bpmrp$ color.  Magnitudes
are from Gaia EDR3; we performed the Gaia DR2 to EDR3 cross-match
using the pre-computed \texttt{gaiaedr3.dr2\_neighbourhood} table
available at the Gaia archive, and required the closest (proper motion
and epoch-corrected) source to be the single match.  Spectral types
are interpolated from the \citet{pecaut_mamajek_2013} table, after
accounting for reddening as described
below.\footnote{\url{http://www.pas.rochester.edu/~emamajek/EEM_dwarf_UBVIJHK_colors_Teff.txt},
version \texttt{2021/03/02}.} Stars in the core appear consistent with having a
fixed age and metallicity, and varying mass.  The halo stars show a
similar sequence, though with greater scatter.  One possible
explanation for the additional scatter is that the halo is more
contaminated by field stars.  For instance, $\approx5$ red giants in
the halo must be field interlopers, because their isochronally implied
ages would be inconsistent with that of the cluster.

Another possible explanation for scatter in the halo's HR diagram
is differential reddening across different sightlines.  The
reported halo spans 20$^\circ$ on-sky, and varies in position from
about $b=-12^\circ$ to $b=-20^\circ$, with the stars closest to the
galactic plane also being further from the Sun by up to 300\,pc
(Figure~\ref{fig:gaia6d}).  An HR diagram binned by galactic latitude
did show some minimal evidence for differential reddening, and so we
expect that both field star contamination and differential reddening
could play a role in the larger scatter of the halo relative to the
core.  A third possibility that we have not explored is whether the
binary fraction could also differ between the different regions of the
cluster.

In the right panel of Figure~\ref{fig:hr}, we compare the observed
Gaia EDR3 photometry with PARSEC isochrones
\citep{bressan_parsec_2012,chen_improving_2014,chen_parsec_2015,marigo_new_2017}.
We used the web
interface\footnote{\url{http://stev.oapd.inaf.it/cgi-bin/cmd},
\texttt{2021/02/26} \texttt{CMD3.4}, \texttt{YBC} bolometric
corrections as in \citet{chen_2019_YBC}.} to interpolate these
isochrones at $\log_{10}(t/\mathrm{yr})=\{8, 8.25, 8.5\}$.

To determine the reddening correction across \cn, we queried the 2MASS
DUST service\footnote{
  \url{http://irsa.ipac.caltech.edu/applications/DUST}; query
  performed using the \texttt{astrobase.services.dust} module
  \citep{bhatti_astrobase_2018}.  } and retrieved the total
line-of-sight extinction parameters at the positions of each \cn\ 
member.  The mean and standard deviation of the
$E(\mathrm{B}-\mathrm{V})$ values for the \citet{schlegel_maps_1998}
and \citet{schlafly_measuring_2011} maps were consistent, so we
adopted the average from \citet{schlegel_maps_1998}:
$E(\mathrm{B}-\mathrm{V})_{\rm S}=0.206\pm0.039$.
\citet{bonifacio_search_2000} found however that the
\citet{schlegel_maps_1998} maps overestimate the reddening values when
the color excess exceeds about 0.10\,mag. We therefore applied the
correction proposed by \citet{bonifacio_search_2000}, and additionally
corrected for the mean cluster distance by a factor
$\exp(-|d\sin(b)|/h)$ where $d$ is the average cluster distance,
$b$ is the average galactic latitude, and $h$ is the scale height of
the galactic disk, assumed to be 125\,pc.  This yielded our adopted
$E(\mathrm{B}-\mathrm{V})=0.102\pm0.019$, where the uncertainty
reflects the standard deviation of the individual
\citet{schlegel_maps_1998} values.  Finally, to convert to Gaia
colors we used the calibration of \citet{stassun_TIC8_2019}, namely
$E(\bpmrp)=1.31 E(\mathrm{B}-\mathrm{V})$ and
$A_G=2.72 E(\mathrm{B}-\mathrm{V})$.  This yielded
$E(\bpmrp)=0.134\pm0.025$, which is used in the
plots.  To redden the isochrones, we assumed $R_V=3.1$, and applied
the \citet{odonnell_1994} SED-dependent extinction law star-by-star
through the PARSEC interface. 

For the metallicity, we considered a range of super and sub-solar metallicities
to fit as much of the locus from $0.5<\bpmrp<1.5$ as possible, and settled by
eye on the slightly super-solar $[M/H]=0.06$ \citep{cummings_2011_li_iron}.
Sub-solar metallicities led to model predictions too blue along the main
sequence by $\approx$0.1\,mag.  Our adopted metallicity agrees with the
spectroscopic metallicity from \citet[][Sec~4.4.4]{cummings_2011_li_iron}, and
with the iron abundance determined by the Gaia-ESO team
\citep{baratella_gaiaeso_2020}. The latter result represented an up-revision
from an earlier sub-solar metallicity \citep{randich_gaiaeso_2018}.  We caution
though that other sub-solar metallicities have been reported
\citep{bailey_rv_2018}, and note that the cluster metallicity and reddening are
degenerate; if the reddening is lowered, the implied metallicity will increase.
While a detailed redetermination of these parameters from the Gaia photometry
is beyond our scope, our adopted values for both the reddening and metallicity
are within the range of previously reported values in the literature.

Overall, the data and models agree for masses above
$\approx$0.45\,$M_\odot$.  Below this mass, the data and models
diverge at colors redder than $\bpmrp\approx2.2$, in
the sense that the model isochrones have lower luminosities and bluer
colors than the observations.\footnote{The 100 Myr isochrone does
overlap well with the lower main sequence, but it fails to fit the
upper main sequence and red giants.}  The MIST isochrones showed a
comparable disagreement \citep{choi_mesa_2016}.  Proposed explanations
for the discrepancy between the models and observations include
starspots and incomplete molecular line lists \citep[{\it
e.g.},][]{stauffer_why_2003,feiden_magnetic_2013,rajpurohit_effective_2013,mann_spectrothermometry_2013,choi_mesa_2016}.
Ultimately, we adopted the PARSEC isochrones because
\citet{chen_improving_2014} implemented an empirical
temperature-opacity calibration, which leads the PARSEC isochrones to
diverge at slightly lower mass than the MIST models.  Regardless, for
purposes of age-dating the cluster we focus on the main-sequence
turn-off (MSTO), since this is where the models have maximum
sensitivity to age.

\citet{cummings_2018} presented techniques for mitigating some of the
complexities of MSTO age-dating ({\it e.g.}, sparse turnoffs, stellar
rotation, high binarity fractions, and the presence of blue
stragglers).  Combining a $UBV$ color-color analysis with Gaia DR2
cluster memberships, they found MSTO ages for NGC\,2516 ranging from
165 to 195\,Myr, depending on the choice of model isochrone (Y$^2$,
PARSEC, MIST, or SYCLIST).

Our goal is to determine whether the ages of the core and halo are
consistent.  The left panel of Figure~\ref{fig:hr} suggests that
isochronally, they are consistent: stars past the turnoff in both the
halo and core are on the same locus.  The right panel of
Figure~\ref{fig:hr} demonstrates the precision with which the claim
can be made.  The MSTO stars are consistent with the 178 and 316 Myr
models, but are bluer than the 100 Myr model.  The red-giant-branch
(RGB) stars are consistent only with the 178 Myr ($\log_{10} t/{\rm
yr} = 8.25$) model.  Based on the assumption that the width of the
MSTO can be attributed to binary stars, we are most interested in the
blue edge (blue stragglers are, however, a concern).  The blue edge
appears most compatible with the 178 Myr model.  These results are
consistent with the MSTO age range of 165 to 195\,Myr found by
\citet{cummings_2018}, and we refer to their work for the more precise
model-dependent comparison.  Based on Table~3 of
\citet{cummings_2018}, we therefore adopt a MSTO age for \cn\ of
$175\pm35$\,Myr, where our quoted uncertainty is a quadrature sum of
the statistical and systematic components from \cite{cummings_2018}.

\subsection{Rotation from TESS}
\label{subsec:tess}

\begin{figure*}[tp!]
	\begin{center}
		\leavevmode
		\subfloat{
			\includegraphics[width=0.700\textwidth]{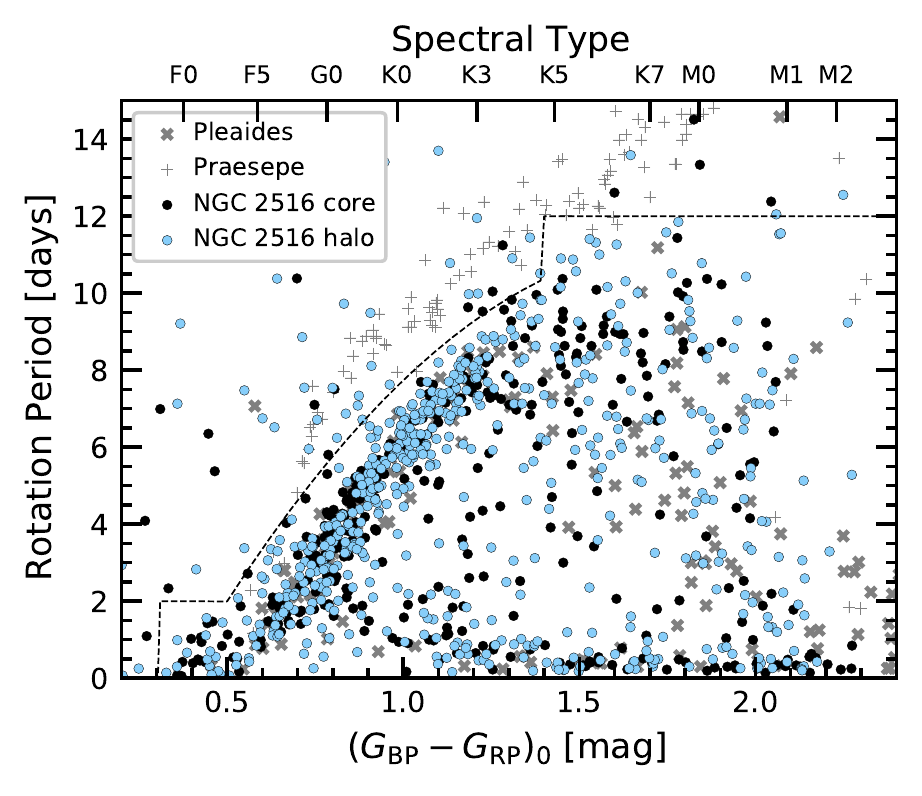}
		}

    \vspace{-0.7cm}
		\subfloat{
			\includegraphics[width=0.700\textwidth]{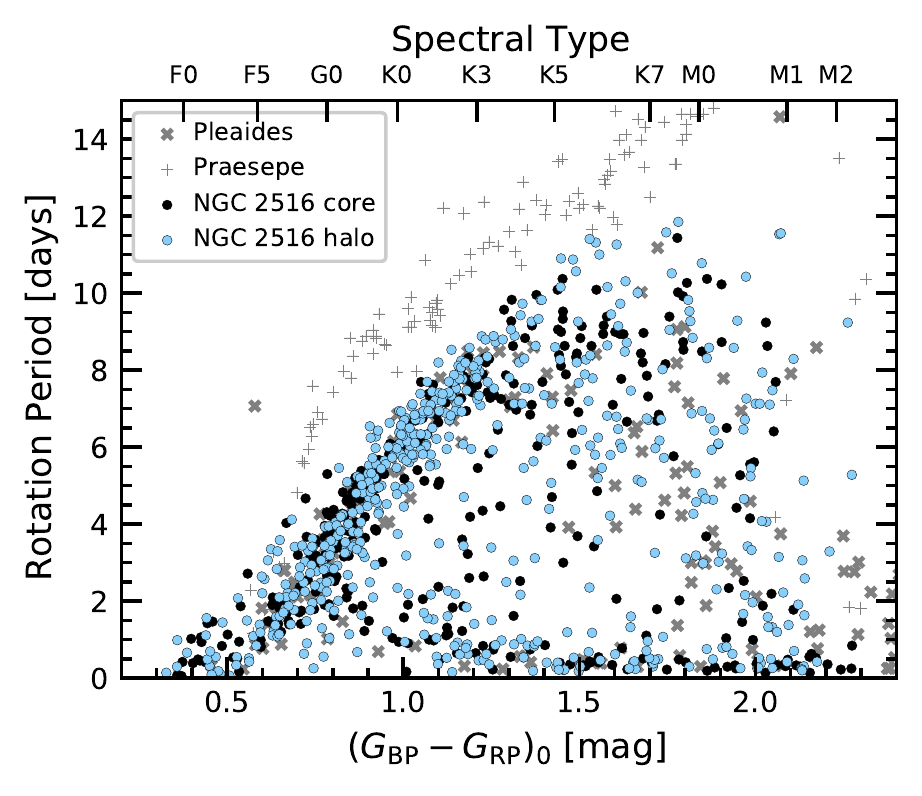}
		}
	\end{center}
	\vspace{-0.7cm}
  \caption{ {\bf TESS rotation periods and dereddened Gaia colors
  across the core and halo of \cn.} Sets $\mathcal{A}$ ({\it top})
  and $\mathcal{B}$ ({\it bottom}) undergo successive stages of
  automated cleaning ; Set
  $\mathcal{B}$ is the same as Set $\mathcal{A}$, after selecting stars
  below the dashed line in the top panel (see Section~\ref{subsubsec:cluster}).  The Pleiades
  \citep[125\,Myr;][]{rebull_rotation_2016a} and Praesepe
  \citep[650\,Myr;][]{douglas_poking_2017} are shown for reference.
  Stars in the core and the halo of NGC\,2516 overlap with the
  Pleiades at $\bpmrpo$$<$1.2.  From $1.2<\bpmrpo<1.7$, some NGC\,2516
  members have longer rotation periods than Pleiades members. Rotation
  periods for a comparison sample of field stars are shown in
  Appendix~\ref{app:compstar}.
  \label{fig:rot}
	}
\end{figure*}
\begin{figure}[t]
	\begin{center}
		\leavevmode
		\includegraphics[width=0.480\textwidth]{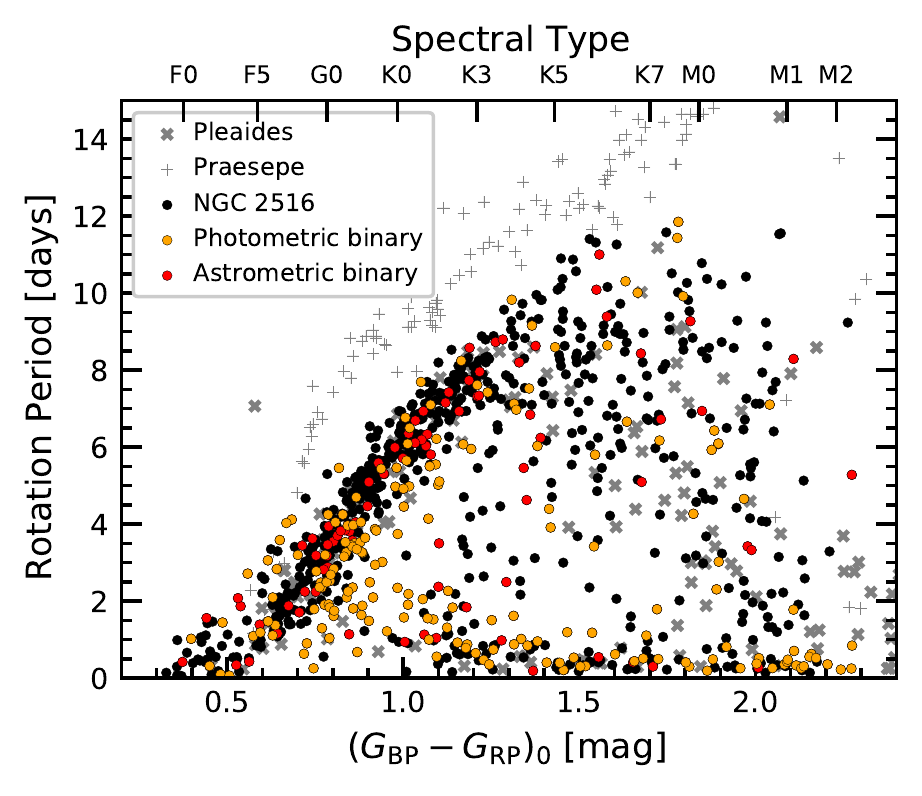}
	\end{center}
	\vspace{-0.7cm}
	\caption{ {\bf Binarity indicators for rotators in \cn.} Set
			$\mathcal{B}$ is shown as in
		Figure~\ref{fig:rot}.  Astrometric binaries ($\mathrm{RUWE}>1.2$;
		11\% of stars) appear in red; photometric binaries ($>$0.3\,mag
		above an empirical isochrone; 20\% of stars) are overplotted in
		orange.  
		Over $0.5<\bpmrpo<1.3$, Set $\mathcal{B}$ has 523 stars.
		Dividing these stars into ``slow'' and ``fast'' sequences by eye,
		the fraction of stars showing signs of binarity in the slow and fast
		subsamples are 27\% (106/289) and
		51\% (68/134), respectively.
		\label{fig:binarity}
	}
\end{figure}

\subsubsection{Methodology}
\label{subsubsec:cluster}

We began our rotation analysis by considering all \nkinematic\
candidate members of \cn.  For each source, we retrieved all available
CDIPS light curves, on a per-sector basis.  This yielded \ncdips\
stars with at least one sector from a CDIPS light curve.  Each of
these stars is brighter than the CDIPS magnitude limit of
$G_\mathrm{RP}=16$\,mag.  \nkinematicrpltsixteen\ of the stars from the \cn\
source list have $G_\mathrm{RP}<16$.  The difference (65 stars) is due to 35 stars
from \citet{meingast_2021} which were not available at the time of the
CDIPS reductions, and 30 stars falling on the TESS chip gaps.  At the
distance of \cn, the $G_\mathrm{RP}=16$\,mag cutoff imposed during the CDIPS
processing corresponds roughly to $\bpmrpo$ of 2.2, or a spectral type
of $\approx$M2V.

We removed systematic trends common to the light curves on each
CCD in each individual sector, and stitched together the resulting
light curves before searching for rotation-induced periodicity.
Details regarding our detrending approach are presented in
Appendix~\ref{app:detrending}.  After detrending, we proceeded with a
few cleaning steps: we masked 0.7 days at the beginning and end of
each spacecraft orbit, and ran a sliding standard-deviation rejection
window over the light curve, which removed any outlying points within
$\pm3\times$MAD of the median in each window.

We then measured the rotation period from the resulting light curve using
the Lomb-Scargle periodogram implementation in
\texttt{astrobase} \citep{lomb_1976,scargle_studies_1982,bhatti_astrobase_2018}.
We used the CDIPS aperture radius that, based on theoretical
expectations, was expected to give the optimal balance between light from
the target star and the sky background \citep{Sullivan_2015}.  This
typically resulted in a circular aperture radius of either 1 or 1.5
pixels.  We recorded the top five periodogram peaks, and their corresponding powers.  Finally, as a check on
crowding, we recorded the number of stars within the aperture with
greater brightness than the target star, and with brightness within 1.25
and 2.5 TESS magnitudes of the target star.

The resulting rotation periods and periodogram powers are reported,
regardless of detection significance, in Table~\ref{tab:maintable}.
To clean these measurements, we designed 
automated cleaning criteria, which yielded sets we denote
$\mathcal{A}$ and $\mathcal{B}$.  For Set
$\mathcal{A}$, we considered light
curves for which the peak Lomb-Scargle periodogram period was below 15
days, the normalized Lomb-Scargle power exceeded 0.08, and for which
no companions with brightness exceeding one-tenth of the target star
were in the aperture ({\it i.e.}, excluding visual binaries
with $\Delta T < 1.25$, according to the Gaia DR2 source catalog).
These limits were chosen after inspecting the light curves visually,
to eliminate low-quality rotation periods while retaining high
completeness.  For Set $\mathcal{A}$, this yielded \nautorotnumerator\
stars, out of \nautorotdenominator\ stars that met the initial crowding requirements.  For Set $\mathcal{B}$, we
additionally required that the Lomb-Scargle period
fell below the boundary drawn in Figure~\ref{fig:rot}, which was
constructed using the Pleiades polynomial from
\citet{curtis_rup147_2020}.
This yielded \nautorotnumeratormatching\ light curves. 
Boolean flags for each set are included in Table~\ref{tab:maintable}.  A
set of field stars was also analyzed for comparison.  The results are
discussed in Appendix~\ref{app:compstar}.
The rationale behind the additional cut in Set $\mathcal{B}$ is
that \cn\ is already known to have few if any stars above the slow
sequence \citep{fritzewski_rotation_2020}.  Stars above the slow sequence
are therefore most likely either not cluster members, or they are cluster
members, but with unresolved binary companions contaminating the rotation
period measurement \citep[{\it
e.g.},][Section~5.1]{stauffer_rotation_2016}.  For the purpose of
assessing which stars are rotationally consistent with being in the cluster,
this is a viable filter; for the purpose of exploring whether any
rotational outliers might be bonafide cluster members, Set $\mathcal{A}$
would be the better cut.

\subsubsection{Rotation-Color Diagrams}

Figure~\ref{fig:rot} shows the resulting rotation periods, with Set
$\mathcal{A}$ shown on top and Set $\mathcal{B}$ on the bottom.  While
Set $\mathcal{A}$ is more complete than Set $\mathcal{B}$, this
completeness comes at the expense of greater contamination.    To facilitate a comparison against the Pleiades and
Praesepe, we used the rotation periods and dereddened colors from Table~5
of \citet{curtis_rup147_2020}, which drew on data from
\citet{rebull_rotation_2016a} and \citet{douglas_k2_2019} respectively.
The first order conclusion is that the Pleiades and \cn\ appear
gyrochronologically coeval for colors spanning $0.5<\bpmrpo<1.2$.  The
implications of this comparison for the age of \cn\ are discussed in
Section~\ref{disc:absage}.

Figure~\ref{fig:binarity} shows the same data, with the points colored by
indicators of binarity.  To assess astrometric binarity, we used the
renormalized unit weight error (RUWE) from Gaia EDR3.  For plotting
purposes, we labeled anything with RUWE exceeding 1.2 as an astrometric
binary (11\% of the overall cluster sample; see {\it e.g.},
\citealt{belokurov_unresolved_2020}).  To assess photometric binarity, we
fitted a spline to Figure~\ref{fig:hr}, fixing the nodes by hand.  We
then labeled any points brighter than 0.3\,mag above the cluster sequence
as photometric binaries.  This included 20\% of the overall cluster
sample.  Many of the resulting candidate binaries, though not all of
them, appear below the slow rotation sequence.  In Set $\mathcal{B}$,
over the color range where the slow and fast sequence are present
($0.5<\bpmrpo<1.3$), there are 523 stars, of which
174 (33\%) show signs of binarity.  Dividing these
523 stars into ``fast'' and ``slow'' sequences by eye,
the fraction of stars showing signs of binarity in the slow and fast
subsamples are 27\% (106/289) and 51\% (68/134), respectively.  We speculate on possible
explanations in Section~\ref{disc:rotn}.

\subsubsection{Comparing Kinematics and Rotation}

\begin{figure*}[tp]
	\begin{center}
		\subfloat{
			\includegraphics[width=0.48\textwidth]{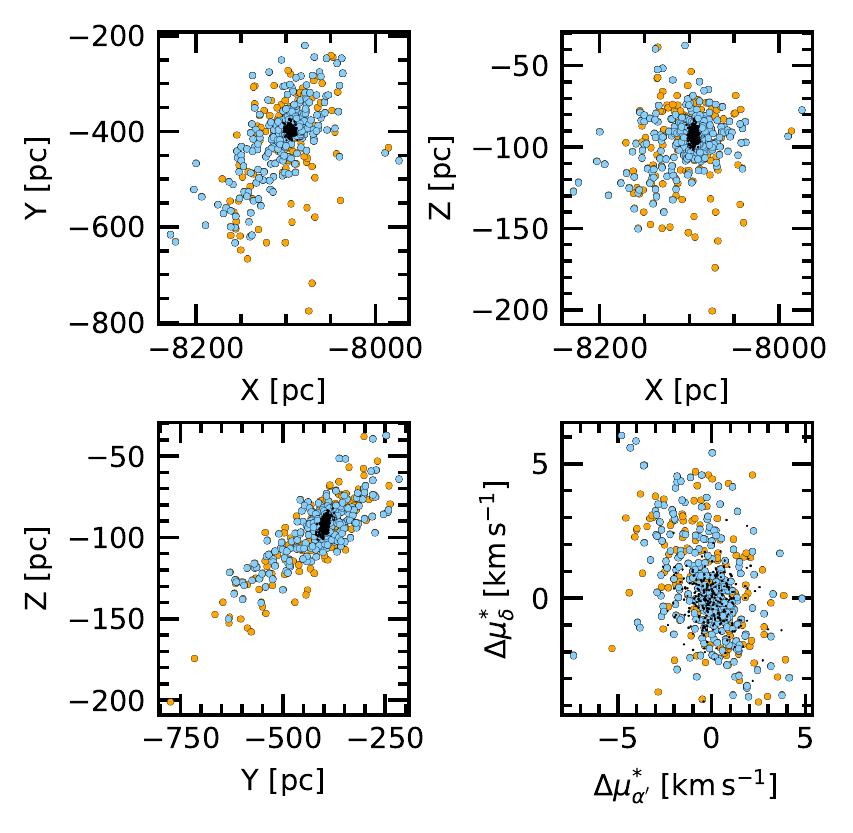}
			\includegraphics[width=0.5\textwidth]{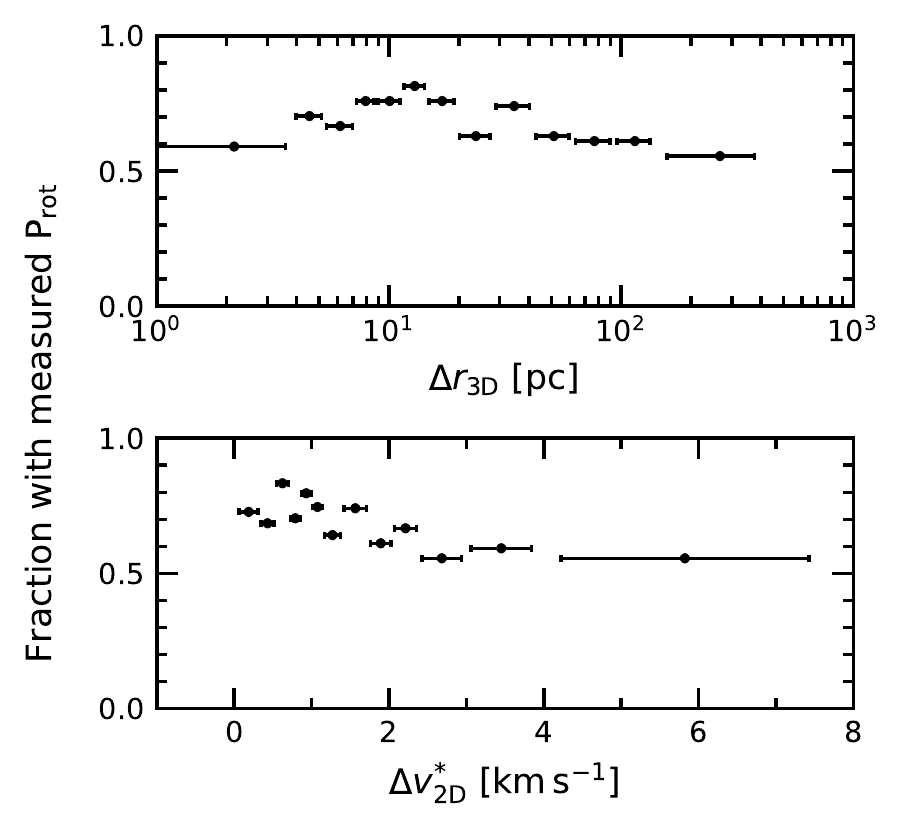}
		}

		\subfloat{
			\includegraphics[width=0.9\textwidth]{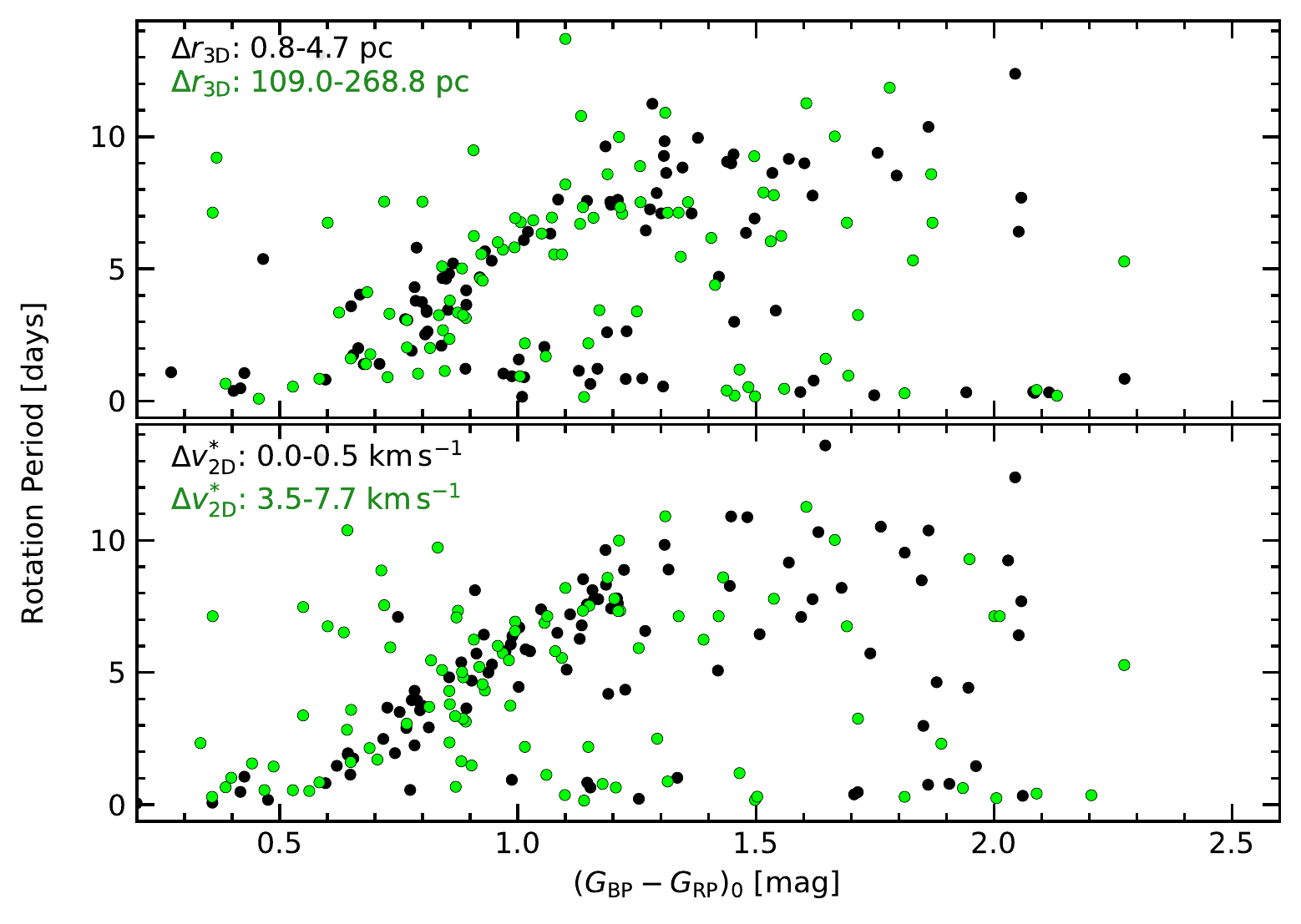}
		}

	\end{center}
	\vspace{-0.7cm}
  \caption{ 
  {\bf \cn\ members exist hundreds of parsecs from the core, and up
  to $\approx$5\kms\ from the core in tangential velocity.} {\it Top
  left}: Cartesian positions (as in Figure~\ref{fig:XYZ}) and
  2D-tangential velocities for halo members.  Stars with detected
  rotation periods in Set~$\mathcal{B}$ are shown in blue;  halo
  members for which rotation periods should have been detectable, but
  were not detected, are shown in orange.  Stars in the core are shown
  as black points.  In
  the {\it top right}, the 
  ratio of detected to expected stars with rotation periods in Set $\mathcal{B}$ is shown
  versus 3D separation from
  the core and 2D tangential velocity difference.  Bin widths are
  chosen to enforce the same number of stars in the denominator
  ($N=54$) per bin.  The {\it bottom panels} show the innermost and
  outermost 100 stars with detected rotation periods in position and
  velocity space (from Set~$\mathcal{A})$.
  \label{fig:physical_x_rotn}
	}
\end{figure*}

The TESS rotation periods provide an independent check on the
Gaia-derived kinematic cluster memberships.  We explored this by
cross-matching the stars with detected TESS rotation periods against
our original target list of \nkinematic\ Gaia members.  

To visualize the results, we again opted for Cartesian coordinates,
and supplemented them by calculating the tangential stellar velocities
relative to the cluster center.  Appendix~\ref{app:vproj} discusses
the projection effect correction required for calculating the
tangential velocities: for a star at position $(\alpha, \delta)$, we
compare the observed proper motion $(\mu_{\alpha'}, \mu_\delta)$ with
what the proper motion at the star's position would have been if the
star were comoving with the core of \cn.  This yields a quantity we
denote $\Delta v^{*}$, per \citetalias{meingast_2021}.  We convert
these proper motion differences to physical units by multiplying by
the measured parallax.

The results are shown in the top two rows of
Figure~\ref{fig:physical_x_rotn}.  Alternative visualizations
in the canoncial space of observed positions, parallaxes, and proper
motions are presented in Appendix~\ref{app:gaia6d_x_rotn}.  In
all these figures, to ensure a fair comparison, we only show
stars with $0.5<\bpmrpo<1.2$ for which our TESS pipeline succeeded in
making light curves.  In other words, the stars in the base sample are
those for which we would have expected to detect a rotation period.
This color range is preferred because the the slow sequence becomes
less defined for spectral types later than $\approx$K4V.  The bins in
the histograms (top-right) were chosen to ensure that the same number
of stars were in each bin.  Rotation periods consistent with a
gyrochronological age of 150\,Myr are detected for
68\% (471/692) of stars reported to be in the cluster.  The average
detection rate inside of 25\,pc, 72\% (290/404), is
slightly higher than outside of 25\,pc, where the detection
rate is 63\% (181/288). Within $\approx$3.5\,pc, the period detection
rate of 59\% (26/44) is
anomalously low, due to crowding and saturation
in the TESS images.

The lower panels of Figure~\ref{fig:physical_x_rotn} give a second
view of how field star contamination affects the outskirts of the
cluster.  Of the outermost 100 stars in position space with detected
rotation periods in Set $\mathcal{A}$, $\approx$8 appear as
outliers above the slow sequence, compared to just a few for the
innermost 100 stars.  For the outermost 100 stars in velocity space,
$\approx$13 appear as outliers.  After accounting for the stars that were
expected to show rotation periods and did not, this suggests field
contaminant rates of $\approx$50\% for the outermost cluster members in
our initial target list.

Despite this level of contamination, the rotation period
measurements show that the halo extends to separations of
$\approx$250\,pc in physical space from the cluster core.  The most
widely separated F2V-K2V stars with rotation periods (blue points in
Figure~\ref{fig:physical_x_rotn}) are separated by $\approx$430\,pc.
The total length of the structure is therefore 400-500\,pc, depending
on which members of the halo are chosen as the ``tips'' on either end.
This agrees with the overall structure of the halo reported by
\citetalias{kounkel_untangling_2019}, with some minor caveats (orange
circles) visible in the top-left panels of
Figure~\ref{fig:physical_x_rotn}.

In projection-corrected tangential velocity space, the fraction of
stars with rotation period detections remains high out to roughly
5\kms.  \citet{meingast_2021} by comparison required a physically
motivated cut in tangential velocity space of 1.5\kms.  Our results
show that at the expense of higher field star contamination rates,
bonafide members can be identified even out at higher velocity
separations.

\subsection{Lithium from Gaia-ESO and GALAH}
\label{subsec:lithium}

\begin{figure*}[t]
	\begin{center}
		\leavevmode
			\includegraphics[width=0.95\textwidth]{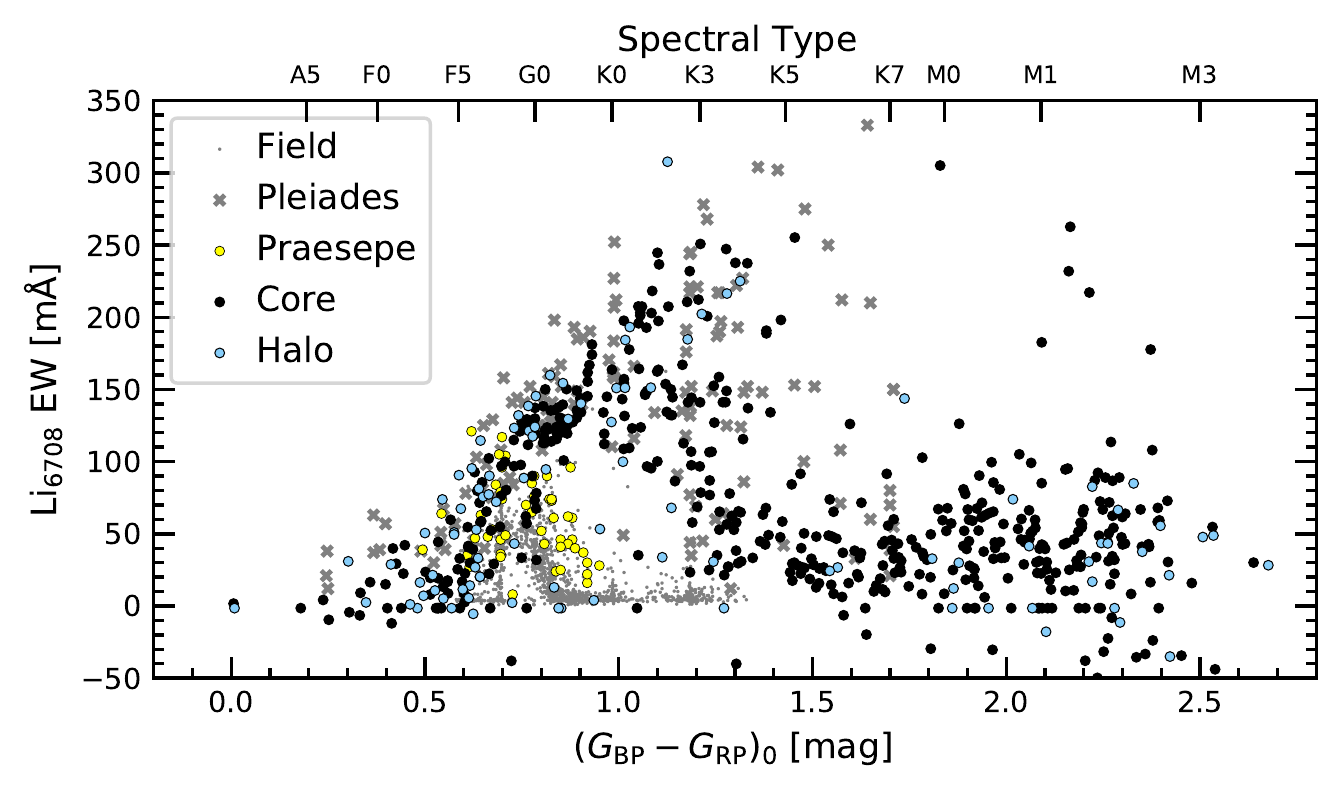}
	\end{center}
	\vspace{-0.700cm}
  \caption{ {\bf Lithium in the core and halo of NGC\,2516.}
  Equivalent width of the 6708\,\AA\ doublet is shown versus
  dereddened color for all candidate \cn\ members with Gaia-ESO
  ($N=459$) or GALAH ($N=107$) spectra available.  The GALAH spectra
  comprise slightly over half of the halo stars, due to the
  non-targeted selection function of that survey.  Gray
  points are field stars with Kepler planets
  \citep{berger_identifying_2018}.  Pleiades stars (125\,Myr)
  with Li detections are from \citet{bouvier_pleiades_lirot_2018};
  Praesepe detections (650\,Myr) are shown in yellow
  \citep{soderblom_praesepe_1993}.  Points with ${\rm
  EW}\approx0\,$m\AA\ are non-detections; all stars 
  later than $\approx$M0 do not have significant lithium.
  The Li measurements for both core and halo stars in \cn\ are
  consistent with a near-Pleiades age.
  \label{fig:lithiumcorehalo}
  }
\end{figure*}

The final approach we took for assessing the youth of the
halo population of \cn\ was an analysis of the Li\,\textsc{I}
6708\,\AA\ doublet.  
For \cn, two spectroscopic datasets seemed important:
Gaia-ESO \citep{gilmore_gaiaeso_2012} and GALAH
\citep{silva_galah_2015}.  At the time of analysis, Gaia-ESO DR4 and
GALAH DR3 were the most relevant
\citep{randich_gaiaeso_2018,buder_galah_2020}.  The target selection
and results from each survey were as follows.

The Gaia-ESO collaboration chose candidate \cn\ members to observe
with GIRAFFE and UVES based on previously reported literature members
and publicly available photometry.  Since the existence of the \cn\
halo was not known at the time of target selection, very few halo
stars are in the sample.  Based on the data,
\citet{randich_gaiaeso_2018} determined stellar parameters (including
lithium equivalent widths and metallicities) for 796 stars that they
considered possible \cn\ members.  Cross-matching these against our
kinematic list of \nkinematic\ candidate members by position and
imposing a 0$\farcs$5 maximum separation limit yielded 459 kinematic
members with available spectra, 417 in the core and 42 in the halo of
\cn.  The lop-sided ratio is due to the Gaia-ESO selection function.
An inspection of the effective temperature {\it vs{.}} lithium
equivalent width based on Table~2 of \citet{randich_gaiaeso_2018}
showed the surprising feature that stars with $T_{\rm eff}\lesssim
4000\,{\rm K}$ often had non-zero equivalent widths, despite the
expectation that their lithium should be fully depleted.  Whether
these measurements were significant was not clear, and so we opted to
download the spectra from the ESO archive to re-measure equivalent
widths.

The GALAH DR3 target selection is discussed by
\citet{buder_galah_2020}.  The relevant aspects for our analysis are
that the survey targeted $12<V<14$ stars at
$\delta<+10^\circ$ in select fields, provided the stars were at least
ten degrees from the galactic plane.  We identified the candidate \cn\ members
for which spectra had been obtained by searching the
\texttt{GALAH\_DR3\_main\_allstar\_v1} catalog, after excluding stars
with the stellar parameter bit flags 1, 2, 3.  This excludes spectra
with unreliable broadening, low S/N, and unreliable wavelength
solutions (see Table~6 of \citealt{buder_galah_2020}).
Of our \nkinematic\ candidate \cn\ members, 107 had spectra in GALAH
DR3.  51 were in the core, and 56 were in the halo.  We
downloaded\footnote{Via \url{datacentral.org.au/services/download},
using the \texttt{sobject\_id} identifiers.} the GALAH DR3 spectra for
all 107 entries.  

%
%

We measured the lithium equivalent widths using the \texttt{specutils}
package \citep{specutils_v1pt1}.  We did this by first shifting the
spectra to the stellar rest frame, and then focused in on a 15\AA\
window centered on the 6707.835\AA\ lithium doublet.  We 
continuum normalized the spectra with a third-degree Chebyshev
series, excluding any regions that showed absorption lines ({\it
e.g.}, Fe\,\textsc{I} is present at 6703.58\,\AA\ and 6705.10\,\AA).
We proceeded by fitting a Gaussian to the continuum normalized
spectra, considering only a $\pm 1\,$\AA\ window centered on the Li
doublet.  The equivalent widths were then evaluated by numerically
integrating the fitted model over the same window.  Our approach
therefore includes the Fe\,6707.44\AA\ blend in the reported Li equivalent
widths---which leads to systematic overestimates of 10 to
15\,m\AA\ \citep[{\it e.g.},][]{bouvier_pleiades_lirot_2018}.
Finally, to derive uncertainties on the EWs, we repeated the procedure
twenty times, but added noise to the spectra, drawn from a normal
distribution with a scale set by the standard deviation of the
continuum.  The reported uncertainties are then drawn from the
$84^{\rm th}$ and $16^{\rm th}$ percentiles of the resulting EW
distribution, with minima imposed at 5\,m\AA\ for FGK stars, and
20\,m\AA\ for M dwarfs. They are included in
Table~\ref{tab:maintable}.  We verified the overall scale of our
results by comparing our measurements with those of
\citet{randich_gaiaeso_2018}.  Most stars with $T_{\rm eff} >
4500\,{\rm K}$ followed a 1-1 relation.  For cooler stars, our EW
measurements are systematically lower, as expected given the depletion
timescales for stars at these ages \citep[{\it
e.g.},][]{soderblom_ages_2014}.  We visually verified from the spectra
that, for the most part, we prefer our measurements.

Figure~\ref{fig:lithiumcorehalo} shows the concatenation of the
Gaia-ESO and GALAH results, with the points colored according to
whether the stars are in the core or the halo of the cluster.  The
GALAH spectra span a color range of $0<\bpmrpo<1.5$, due to the $V<14$
brightness cutoff of the survey.  The overall increase of Li EW from
$0<\bpmrpo<1.0$ is driven by the temperatures in the stellar
chromospheres: hotter stars fully ionize Li out of its ground state
\citep[{\it e.g.}, Figure~4 of][]{soderblom_evolution_1993}.  The
depletion of stars redder than $\bpmrpo\gtrsim1.4$, {\it i.e.}, later
than K4.5V (0.71\,$M_\odot$), is visible relative to the earlier K
dwarfs: these stars have burned their lithium.  For comparison, we
have also plotted the Li EWs measured by
\citet{berger_identifying_2018} for planet-hosting stars in the Kepler
field, and Pleiades members from \citet{bouvier_pleiades_lirot_2018}
(we interpolated from their effective temperatures using the
\citealt{pecaut_mamajek_2013} table).  For clarity, we have omitted the
few upper limits reported by \citet{bouvier_pleiades_lirot_2018}.  The
field star distribution peaks at late F-type stars.  The majority of
kinematically selected stars in the core and the halo show lithium
equivalent widths substantially in excess of these field stars.

\begin{figure*}[t]
	\begin{center}
		\leavevmode
		\includegraphics[width=0.95\textwidth]{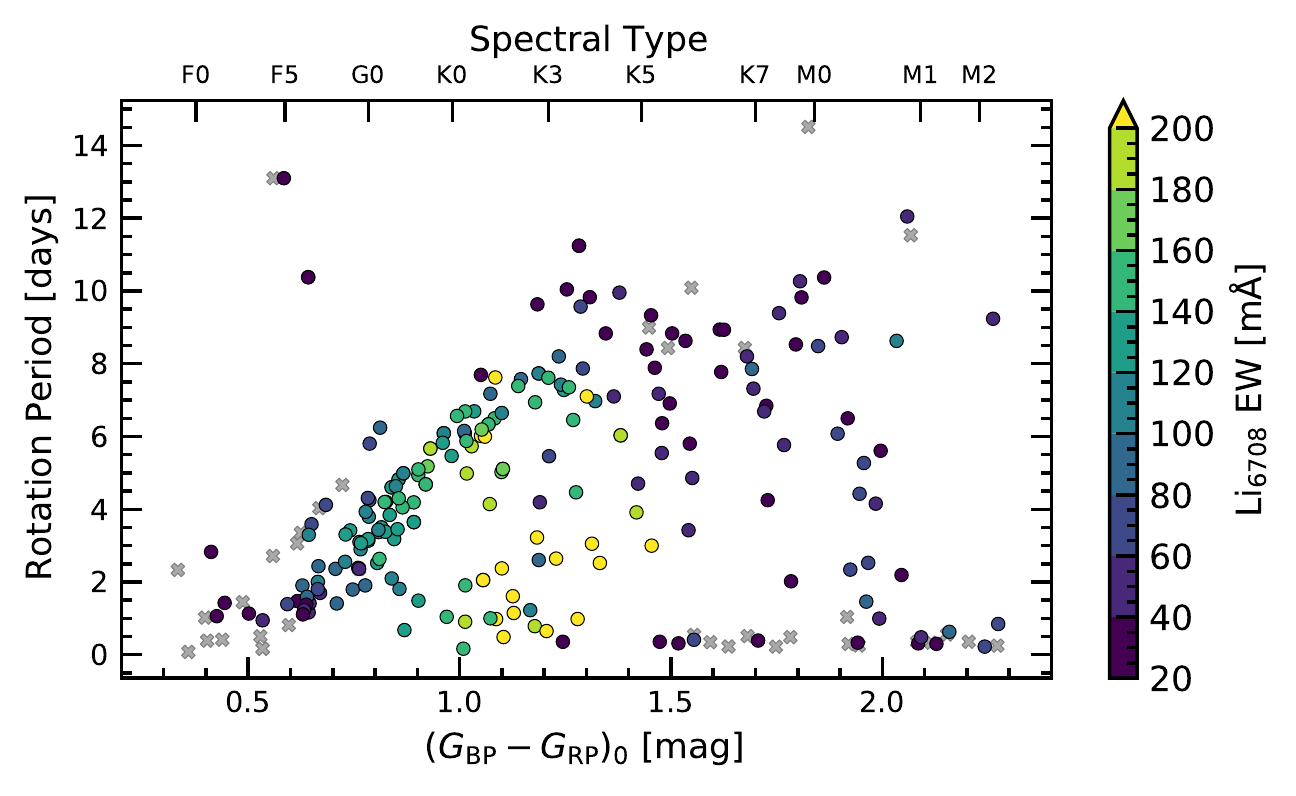}
	\end{center}
	\vspace{-0.7cm}
	\caption{ {\bf Lithium and rotation in NGC\,2516.} Points show all
		candidate \cn\ members with TESS rotation periods (Set $\mathcal{A}$) and Gaia-ESO or
		GALAH spectra, and are colored by equivalent width.  Points with
		reported equivalent widths below 20\,m\AA\ are shown with gray
		crosses.  The lithium-rotation correlation for the K-dwarfs has been
		observed in other young clusters, and is discussed in
		Section~\ref{discussion:lithium}.
		\label{fig:lithiumrot}
	}
\end{figure*}

We can go one step further, and match our sample of kinematically
selected stars with spectra against the TESS rotators.  The result is
shown in Figure~\ref{fig:lithiumrot}. At fixed stellar mass, the rapid
rotators have lithium equivalent widths an order of magnitude larger
than the slow rotators.  This effect is mostly apparent in the K
dwarfs.  Similar trends were noted in the Pleiades over three decades
ago \citep{butler_pleiades_1987}, and are thought to be caused by
differences in lithium abundance
\citep{soderblom_evolution_1993}.  Alternative explanations, such as
differences in line-formation conditions ({\it e.g.,} chromospheric
temperatures, microturbulent velocities, or the presence of starspots)
are incompatible with the available data.  The lithium-rotation
correlation is discussed further below
(Section~\ref{discussion:lithium}).

\section{Discussion}
\label{sec:discussion}

\subsection{How did the halo form?}
\label{subsec:origin}

Any theory for the structure of \cn, the Psc-Eri stream, and the
strings and halos found by \citetalias{kounkel_untangling_2019} and
\citetalias{meingast_2021} needs to explain a few observations.  First
is their aspect ratios.  The halo of \cn, for instance, extends over
at least 500\,pc, despite being only $\approx$25\,pc wide.  Second is
their axial tilts in the plane of the Milky Way -- why their leading
and trailing arms tend to conform to the Galaxy's differential
rotation ({\it e.g.}, Figure~\ref{fig:XYZ} or Figure~13 of
\citetalias{meingast_2021}).  Third is the correlation between the
mean LSR-subtracted cluster velocity and elongation axes (blue arrows
in Figure~\ref{fig:XYZ}).  Such a correlation has also been noted in
Coma Ber by \citet{tang_comaber_2019}; the other halos may show
similar trends.

The most likely explanation seems to be that the observed halo
structures are tidal tails \citep[{\it
e.g.},][]{chumak_tails_2006,krumholz_star_2019}. The idea is that
stars near the cluster's tidal radius escape slowly due to the
galactic tide, and subsequently form leading and lagging arms due to
differential rotation in the Galaxy.  The contraction rate of the
cluster can affect this process.  Whether the exact contraction rate
of \cn\ \citep{healy_stellar_2020}, and the corresponding evaporation
rate of the core are sufficient to explain the formation of the halo
has yet to be assessed. 

A second possibility is that the clusters form in larger and more
dispersed star formation complexes: the stars in the halo need not
have formed in the same ``clump'' as those in the cluster core.  In
other words, the formation environment might be a
giant molecular filament, rather than a giant molecular cloud.  A
sample of such filaments has been collected and analyzed by
\citet{zucker_physical_2018}: if the current $\approx$20:1 aspect
ratio of \cn\ were primordial, its structure would match the aspect
ratios of what they term either ``elongated dense core complexes'', or
``bone candidates''.  A more immediate example is the Orion\,A cloud.
\citet{grosschedl_3d_2018} showed using young stellar objects as
tracers that Orion\,A is 90\,pc long, and that it has a dense ``head''
and a lower density ``tail''.

\begin{figure}[t]
	\begin{center}
		\leavevmode
		\includegraphics[width=0.5\textwidth]{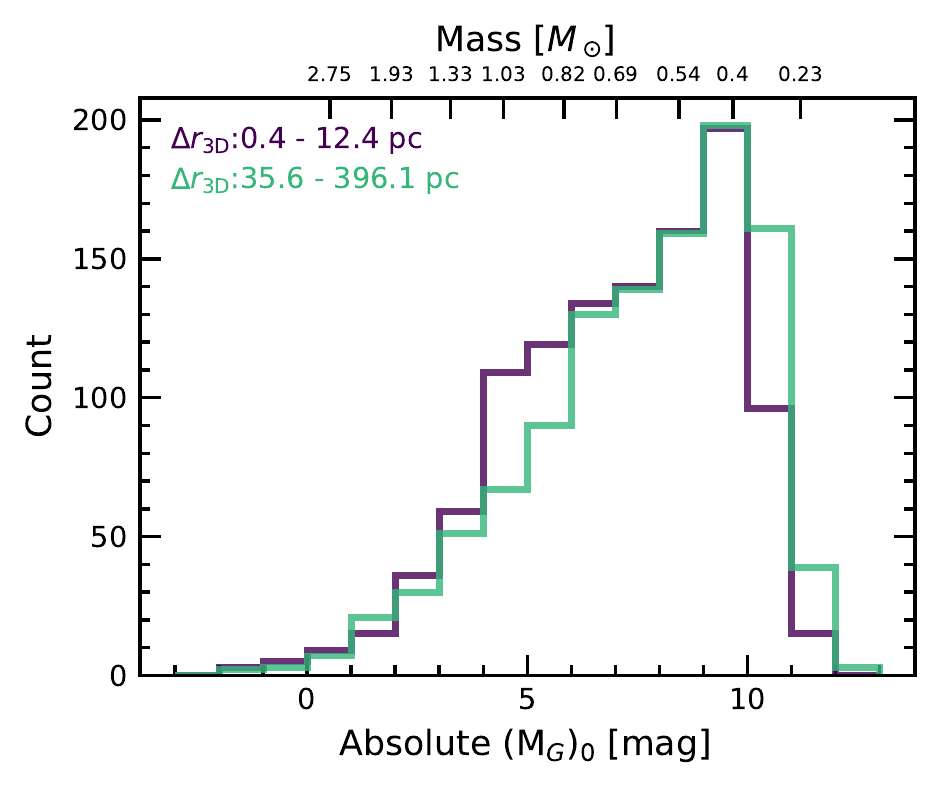}
	\end{center}
	\vspace{-0.7cm}
  \caption{ {\bf Masses and absolute magnitudes for the innermost and
  outermost stars in \cn.} The \nkinematic\ candidate \cn\ members
  were split into three groups according to their distance from the
  cluster center.  The inner and outermost groups, each containing
  1,099 stars, are counted in 1\,mag bins.  The mean star in the
  innermost sample has a 10\% higher mass than the mean star in the
  outermost sample, consistent with expectations for tidal tails.
  \label{fig:massfunction}
	}
\end{figure}

Any differences in the stellar mass function between the core and halo
could be informative, since the tidal tail explanation predicts that
the mean tail star should be less massive than the mean core
star \citep[{\it e.g.},][]{chumak_tails_2006}.
Figure~\ref{fig:massfunction} shows the histogram of $A_G$-corrected
absolute magnitudes, with the corresponding
\citet{pecaut_mamajek_2013} mass estimates.  The stars closest to the
cluster center do seem to have a greater proportion of $\gtrsim$0.8$M_\odot$
stars, while the outermost stars have a greater proportion of
$\lesssim$0.4$M_\odot$ stars.  The mean star in the inner separation
bin has $({\rm M}_G)_0=7.08$ ($\approx$0.69$M_\odot$), while in the
outer bin the mean star has $({\rm M}_G)_0=7.61$
($\approx$0.63$M_\odot$).  While the difference does appear
significant given the number of stars involved, the theoretical
implications are not entirely clear, since the ``molecular filament''
explanation might be able to accommodate mass differences between the
core and tail through density gradients across the initial filament.
Observational incompleteness at the low-mass end seems unlikely to
alter the result, since the innermost and outermost regions are 
subject to similar selection effects.

One possible approach for distinguishing the two explanations could be
to observationally establish an age-size correlation.  If the
$\approx$100\,Myr halos are produced primarily through evaporation and
expansion, then the younger clusters should be significantly smaller
\citep{chumak_tails_2006,chumak_realOCs_2006}.  A theoretical approach
along complementary lines would be to perform a dynamical analysis of
whether filaments like Orion\,A, or those in the
\citet{zucker_physical_2018} sample are at all expected to evolve into
structures resembling \cn.

\subsection{Stellar rotation at 100-200\,Myr}
\label{disc:rotn}

\begin{figure*}[t]
	\begin{center}
		\leavevmode
		\includegraphics[width=0.95\textwidth]{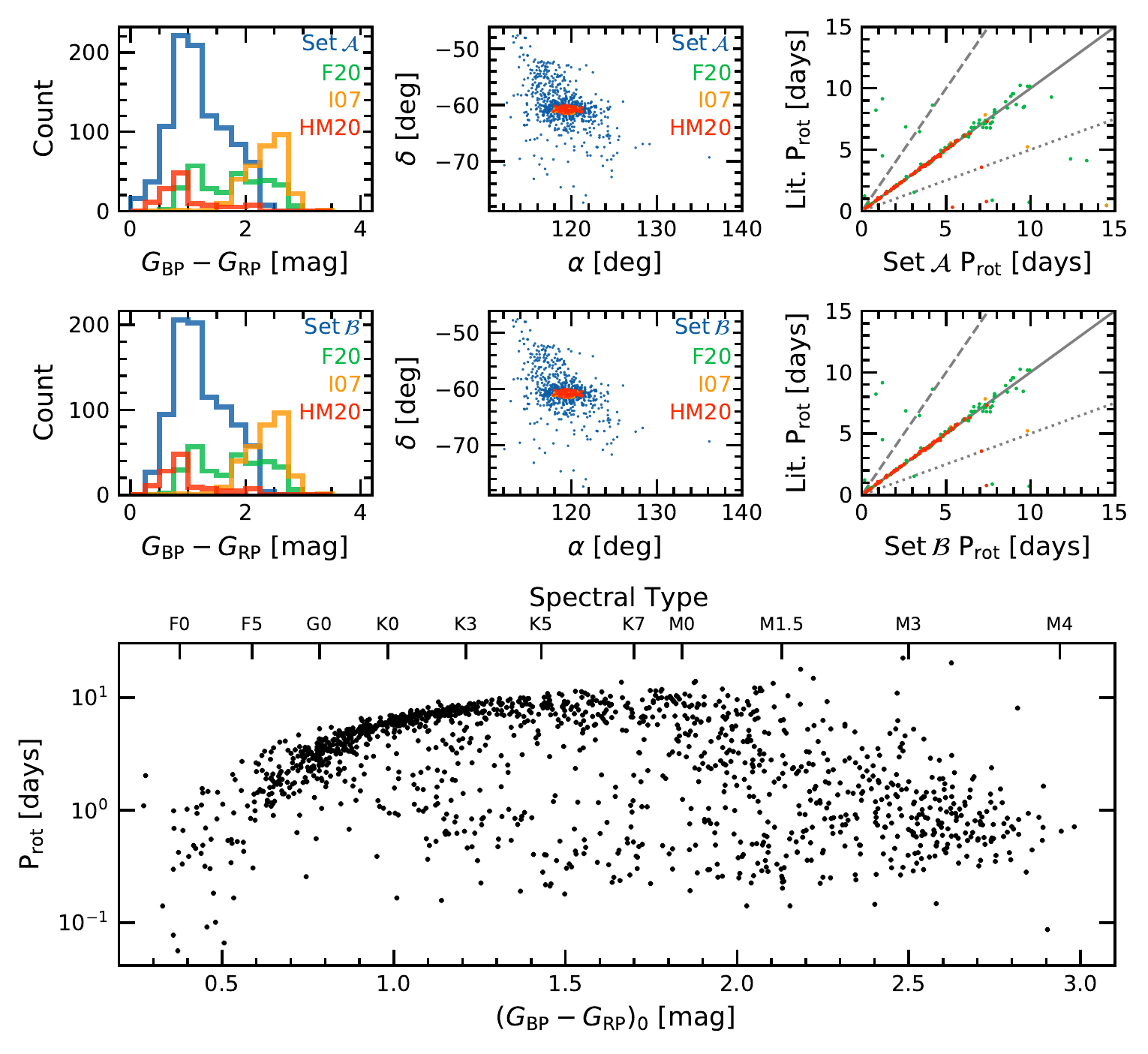}
	\end{center}
	\vspace{-0.7cm}
  \caption{ {\bf Comparison against previous analyses of stellar
  rotation in \cn.} 
  The bottom panel is a concatenation of Set $\mathcal{B}$ with the
  results of \citet{Irwin_NGC2516_2007}
  (I07), \citet{fritzewski_rotation_2020} (F20), and
  \citet{healy_stellar_2020} (HM20).
  I07, F20, and HM20 analyzed stars in the
  core of \cn.
  \label{fig:litcomp}
	}
\end{figure*}

From their rotation period analysis of \cn,
\citet{fritzewski_rotation_2020} demonstrated that between 100 and
200\,Myr, young clusters (\cn, Blanco\,1, Psc-Eri, M\,35, M\,50, and
the Pleiades) all show similar rotation period distributions.  The FGK
stars show a ``slow sequence'', a population of ``rapid rotators'',
and in some cases stars in the ``void'' between the two populations
({\it e.g.,}, Psc-Eri and \cn\ show more stars in the void than the
Pleiades).  The late M dwarfs with detected rotation periods tend to
rotate rapidly ($<3$\,days), though there may also be late M dwarfs
with slow ($>20$\,days) rotation periods
\citep{stauffer_rotation_2016}.  Although these late M dwarfs are
beyond the magnitude limit of our analysis, they have been explored in
depth by \citet{Irwin_NGC2516_2007} and
\citet{fritzewski_rotation_2020}.  A comparison of our results is
shown in Figure~\ref{fig:litcomp}.  The bottom panel of this plot
includes, in order of precedence: rotation periods from our analysis,
from ``Class 1 and
2'' \citet{fritzewski_rotation_2020} sources, from
Table 1 of \citet{healy_stellar_2020}, and
from \citet{Irwin_NGC2516_2007}.  The main
contribution of our study beyond these previous analyses is our
sensitivity across both the core and halo of \cn, and a corresponding
expansion in the number of stars that can be analyzed.  In the
following section, we therefore focus on the origins of the features
in the rotation-color diagram, and whether our expanded sample can
provide any new insights.

The existence of a slow sequence is the main basis of gyrochronology
\citep[{\it e.g.},][]{barnes_rotational_2003}.  While the 
spindown relation ($P_{\rm rot}\sim t^{1/2}$) proposed by
\citet{skumanich_time_1972} is not an accurate enough description, the
general idea is correct, and the first-order effect can be understood
through magnetic braking \citep{weber_angular_1967}.

The other features of the rotation-color diagram are indications that
magnetic braking is not always the only important effect.  The
question of why the rapid rotators exist, for instance, requires
additional physics.  Models that vary core-envelope decoupling
timescales and pre-MS disk lifetimes can reproduce each type of
population
\citep{Irwin_NGC2516_2007,gallet_improved_2013,gallet_improved_2015};
the question then shifts to why these timescales should differ for
stars that are otherwise identical.  Another possibility
is that proposed by \citet{matt_mass-dependence_2015}: the
saturation of the magnetic dynamo yields ${\rm d}P/{\rm d}t \sim P$ when $P >
P_{\rm sat}$, ${\rm d}P/{\rm d}t \sim P^{-1}$ for $P < P_{\rm sat}$, and a
maximal spin-down rate in the intermediate regime.  This could
naturally produce the period bimodality observed at ages of
$\approx$100\,Myr (though it fails to explain the widening of the
distribution observed at later times
\citealt{godoyrivera_stellar_2021}).  Other significant processes
could be external to the star entirely: \citet{qureshi_signature_2018}
for instance have reported that mergers between giant planets and the
star could explain a non-neglible fraction of the rapid rotators.

A separate hypothesis that we can test using the existing data is the
idea that rapid rotation can be explained through binarity.
Correlations between rapid rotation and binarity in both young
clusters and the field have previously been noted
\citep{meibom_effect_2007,stauffer_rotation_2016,simonian_rapid_2019,gillen_ngts_2020}.
Related effects could include tidal synchronization for the shortest
period binaries, or alternatively pre-MS magnetic locking between the
disk and star \citep[{\it
e.g.},][]{koenigl_disk_1991,long_locking_2005}.

The lower panels of Figure~\ref{fig:rot} show our attempt at identifying
unresolved binaries in the \cn\ rotation sample.  Binaries resolved by
Gaia were already excluded in our initial definitions of the samples.  In
\cn\, we see that the fraction of stars showing signs of binarity in the
slow and fast rotation subsamples are 26\% (106/289) and 51\% (68/134); the fast rotators show a
preference for binarity.  This correlation is in line with the earlier
findings noted above.  While many rapid rotators in \cn\
(66\%) have no evidence for being binaries, the cuts we
used to select binaries (${\rm RUWE}>1.2$; photometric excess
$>$0.3\,mag) leave open a significant fraction of parameter space.  A
0.3\,mag flux excess, for instance, corresponds to mass ratios $M_2/M_1
\gtrsim 0.70$, assuming the luminosity $L$ scales as $M^{3.5}$. 

Comparing the explanations of disk-locking against tidal
synchronization, the population statistics seem to rule out 
tidal synchronization.  A quarter of the \cn\ members
are rapidly rotating.  In comparison, in the field half of Sun-like
stars are binaries, and only $\approx$9\% of these binaries have
periods below 100\,days \citep{raghavan_survey_2010}.  If we assume
that all binaries with sub-100\,day periods become tidally
synchronized, then we can only explain a rapid rotator occurrence rate
of $\approx$5\%.  Elevated binarity fractions in pre-MS stars likely
do not change this picture \citep[see Section~4.4
of][]{duchene_stellar_2013}.

A separate effect is that on the slow rotation sequence,
Figure~\ref{fig:rot} shows that the binary stars appear to be either
preferentially redder, or to have faster rotation periods than the
single stars.  One likely explanation for this could be that the
unresolved binaries have a component contributing additional red light
to the system, skewing the color measurement of the primary.  Whether
any physical effects could be at play remains a question for future
work.

\subsection{The lithium-rotation correlation}
\label{discussion:lithium}

The lithium-rotation correlation (Figure~\ref{fig:lithiumrot}) carries
over beyond just equivalent widths, and to the actual abundance of
lithium in the photosphere \citep{soderblom_evolution_1993}.  Non-LTE
corrections do not change the overall relationships between Li
abundance and stellar temperature or rotation
\citep{carlsson_1994,lind_departures_2009}.  The correlation has been
seen in the Pleiades, the Psc-Eri stream, and M\,35, among other
clusters
\citep{bouvier_pleiades_lirot_2018,arancibia_2020,jeffries_m35_li_2020,hawkins_2020}.
In \cn, \citet{jeffries_rotation_1998} tentatively observed it as
well, when analyzing 24 stars in the core of \cn.  Our concatenation
of the Gaia-ESO and GALAH-DR3 spectra represents a significant
expanasion in volume and color range from this earlier analysis.

What causes the correlation between lithium abundance and rotation at
fixed stellar mass?  Hints likely lie in shared correlations between lithium
abundance, rotation rate, radius inflation, internal mixing, and
magnetic field strength
\citep{chabrier_evolution_2007,somers_measurement_2017,jeffries_m35_li_2020}.
For instance, one interpretation of the lithium-rotation correlation
based on pure hydrodynamics is that the rapid rotation leads to less
efficient convective penetration (``overshoot''), which lowers the mixing
efficiency at the convective-radiative boundary
\citep{baraffe_lithium_2017}.  The magnetic field itself may even be
sufficient to inhibit the convection \citep{ventura_Li_B_1998}.
Alternatively, the star's temperature and density profiles could be at fault: rapid rotators
have the strongest magnetic fields and the most spotted surfaces.
These spots lower the photospheric flux, and drive the stellar radius
to expand while lowering the core temperatures, which in turn slows
the rate of lithium-destroying proton capture reactions
\citep{feiden_magnetic_2013,somers_rotation_2015}.  

None of the internal processes noted above answer the initial
question of why some of the G and K dwarfs rotate rapidly to begin
with.  One clue could be that the rapid rotators tend to be in
photometric or astrometric binaries.  The majority of such binaries
have wide ($a\gtrsim200$\,au) or intermediate (0.5 - 100\,au)
separations \citep{raghavan_survey_2010}.

It is expected that the circumstellar disks in these binaries have
different properties than those of single stars, due to {\it e.g.}, gap
formation in the circumbinary disk, disk truncation, and faster disk
clearing \citep{artymowicz_dynamics_1994,moe_impact_2020}.  It would
therefore be reasonable to assume that disk lifetimes in the binary
systems are shorter than in single star systems.
\citet{eggenberger_impact_2012} showed that longer disk lifetimes
should lead to an extended phase of disk-locking, which leads to a
greater amount of differential rotation generated in the radiative
zone, more efficient shear mixing, and consequently a lower photospheric
lithium abundance.  This provides a plausible explanation for the
overall trend that the rapid rotators are often in unresolved
binaries, and that they are also lithium rich.  A careful
observational analysis including an assessment of the completeness to
different separations and mass ratios of binaries would be useful to
clarify whether or not all of the rapid rotators are in binaries.  If
not, this scenario would be falsified.

\subsection{The age of NGC\,2516}
\label{disc:absage}

\paragraph{Age from Gyrochronology}
Comparing the slow sequence of the Pleiades and \cn\ more closely, the
top row of Figure~\ref{fig:rot} shows that the two sequences overlap
from $0.5<\bpmrpo<1.2$.  At redder colors, $1.2<\bpmrpo<1.7$, the
dispersion in rotation periods increases.  The maximum rotation
periods seen in \cn\ 
extend up to $\approx$11 days, rather than the $\approx$8.5 day upper
limit seen in the Pleiades.  This is consistent with \cn\ being
slightly older than the Pleiades.

Fitting a model to substantiate the claim that \cn\ is
gyrochronologically older than the Pleiades \citep[{\it e.g.},][]{mamajek_improved_2008,angus_toward_2019,spada_competing_2020}
is, unfortunately, an exercise in tautology.  Gyrochronological models
are empirically calibrated against the Pleiades and Praesepe.  At
$\bpmrpo=1.35$, we observe rotation periods in \cn\ ranging from 7 to
10 days.  The 7 day rotation periods are consistent with the Pleiades;
the 10 day rotation periods are not.  Fitting formulae from the
previously cited studies give ages for a star with $\{\bpmrpo=1.35,
(\mathrm{B}-\mathrm{V})_0=1.10, P=7\,{\rm day}\}$ of 204 Myr, 316 Myr,
and 107 Myr respectively.  Given a Pleiades age of $125\pm20$\,Myr (an
average of MSTO and rotation-corrected lithium-depletion results; see
{\it e.g.} \citealt{stauffer_keck_1998},
\citealt{soderblom_ages_2014}, and \citealt{cummings_2018}), only the
\citet{spada_competing_2020} model has an absolute scale that seems to
be well-calibrated at these ages and colors.  For an 8-day rotation
period of a star with the same color, this model quotes a 193 Myr age;
at 9 days, 388 Myr.  Given this degree of model uncertainty, we prefer
the relative statement that \cn\ appears slightly gyrochronologically
older than the Pleiades, and much younger than Praesepe.  A
gyrochronological age estimate for \cn\ between 150 and 200\,Myr would
therefore appear reasonable.  Much older, and the issue of why the F
and G dwarfs do not spin down becomes pressing.  We caution however
that the spindown rate of FGK stars between 100 and 300 Myr has not
yet been empirically calibrated, and if ``stalling'' were to occur, it
could affect our age assessment \citep[see][]{curtis_rup147_2020}.

\paragraph{Age from Lithium Depletion}
The lithium depletion boundary for \cn\ has to our knowledge not been
measured.  Given the distance to the cluster, this is not surprising:
at $\sim$150\,Myr, the LDB is expected to occur at a stellar mass of
$\approx$0.088\,$M_\odot$ \citep{soderblom_ages_2014}.  This would
correspond to $\bpmrpo\approx4.7$, well beyond the limits of available
cluster membership lists and spectroscopy.

An easier {\it relative} measurement to make of lithium depletion is
from the EW-color diagram (Figure~\ref{fig:lithiumcorehalo}).  The
\cn\ and Pleiades sequences in this diagram appear indistinguishable.
A relative lithium-based age estimate for \cn\ would therefore be
``Pleiades-age''.  The uncertainties on this estimate are large
because lithium depletion timescales are long for Sun-like stars past
100\,Myr.  For example,
\cn\ FGK stars are on average more depleted than stars in
IC\,2602 ($\approx40\,{\rm Myr}$), but only marginally so \citep{soderblom_ages_2014}.  On the
older end though, Figure~\ref{fig:lithiumcorehalo} does show that
relative to Praesepe ($\approx$650\,Myr), the lithium EWs of \cn\ are
significantly enhanced.  Given these comparisons, a plausible range of
ages based on lithium for \cn\ would therefore be between $\approx$100
and 200\,Myr.

\paragraph{Adopted Age}
As noted in Section~\ref{subsec:hr}, a color-color and color-magnitude analysis of the
main sequence turn-off (MSTO) by \citet{cummings_2018} found that \cn\
is 40-60\,Myr older than the Pleiades.  Given a Pleiades age of
$125\pm20$\,Myr, this yields a MSTO age for \cn\ of $175\pm35$\,Myr.
The gyrochronological age we have argued is likely older than the
Pleiades, {\it i.e.}, in the range of 150 to 200\,Myr.  The lithium
age based on the depletion of the G and K dwarfs is more uncertain,
but is consistent with the Pleiades and could be a bit older
($\approx$100 to 200\,Myr).  Averaging these three different
indicators gives an age for \cn\ of $167\pm20$\,Myr, if we set the
uncertainties to match the absolute Pleiades age uncertainty of $\pm
20\,$Myr \citep{soderblom_ages_2014}.  Whether this average age is at
all meaningful, we leave the reader to judge.

\section{Conclusion}
\label{sec:conclusion}

The combination of astrometry from Gaia, photometry from TESS, and
spectroscopy from GALAH and Gaia-ESO has clarified a few things about
the structure of \cn.
\begin{itemize}
  \item {\it Over-densities observed in the Gaia 3D positions and 2D
    velocities revealed a halo of stars spanning $\approx$500\,pc
    in the plane of the galaxy.} The earliest reference to this halo
    that we can find is by \citet{kounkel_untangling_2019}.  The halo
    is more precisely described as a leading and trailing tail
    (Figure~\ref{fig:XYZ}), with the leading edge angled toward the
    galactic center, relative to the cluster's orbit in the Galaxy.
    This is consistent with the direction and amplitude of the Milky
    Way's differential rotation
    ($\approx$-0.2\,deg\,Myr$^{-1}$\,kpc$^{-1}$), and similar
    halos/tails have been observed around $\approx$10 other nearby
    open clusters \citep{meingast_2021}.
  \item {\it Isochronal, rotational, and lithium dating show that the
    halo of \cn\ is coeval with its core.} In short, all the data are
    consistent with the halo being real.  The Gaia EDR3 data
    show a main-sequence turnoff that suggest an isochronal
    age of 150\,Myr for both the core and halo (Figure~\ref{fig:hr}).
    The faintest M dwarfs in the core and halo are also brighter than
    field stars of the same color.  Gyrochronally, stars in
    \cn\ spanning $0.5<\bpmrpo<1.2$ (spectral types F2V-K3V) overlap
    with the slow sequence of the Pleiades (Figure~\ref{fig:rot}).  At
    redder colors, from $1.2<\bpmrpo<1.7$ (spectral types K3V-K6V), there is a larger
    dispersion in rotation period.  The upper envelope of the rotation
    period distribution at these redder colors extends to longer
    rotation periods than in the Pleiades (11 {\it vs.}\ 8 day
    rotation periods).  This is consistent with \cn\ being slightly
    older than the Pleiades ({\it i.e.}, $\approx$150\,Myr).  The
    lithium equivalent widths of the kinematic cluster members
    (Figure~\ref{fig:lithiumcorehalo}) are consistent with this
    assessment.
  \item {\it Bonafide \cn\ members exist out to $\approx$250\,pc in
    separation and out to $\approx$5\kms\ in tangential velocity
    separation from the cluster core.}
    Figure~\ref{fig:physical_x_rotn} shows the overlap between
    kinematic and rotational cluster members used to make this
    assessment.
   	The majority of the halo members do however exist within
   	tangential velocity separations of $\approx$2\kms.
  \item {\it The field star contamination rate increases at larger
    physical and velocity separations from the cluster core.} The
    right and lower panels of Figure~\ref{fig:physical_x_rotn}
    quantify the statement: $\approx$70\% of the innermost
    candidate members have detected rotation periods consistent with a
    gyrochronological age near that of the Pleiades. At physical
    separations of $\approx$200\,pc, this fraction decreases to
    $\approx$50\%.  While the latter contamination fraction may seem high, a comparison sample of field stars (Appendix~\ref{app:compstar}) has a detection rate three times lower, and a  distribution in the period-color plane inconsistent with that of a population with a single age.
   \item {\it The average star in the outskirts has lower mass (by $\approx$10\%)
   than the average star in the cluster center.}
   This is shown in Figure~\ref{fig:massfunction}, and discussed in
   Section~\ref{subsec:origin}.
   While this is consistent with expectations for a
   ``tidal tail'' origin of the halo, the possibility of a primordial halo
   also merits exploration.
   \item {\it The rapidly rotating K dwarfs show elevated lithium abundances, and
     elevated binarity fractions}.  Each trend has been noted in
     comparable stellar populations ({\it e.g.,}
     \citealt{soderblom_evolution_1993};
     \citealt{meibom_effect_2007};
     \citealt{jeffries_m35_li_2020}; \citealt{gillen_ngts_2020}).  The
     rough orders of magnitude of the effects are $\approx$1\,dex in
     lithium abundance for a factor of 10 in rotation period, and a
     factor of $\gtrsim2$ enhancement in binarity fraction for fast
     {\it vs{.}} slow rotators.  
     Section~\ref{discussion:lithium} discusses one plausible
     connection between the two observations: the disk-locking
     phase may be truncated in binary systems, which could free the
     stars to spin up as they contract on the pre-MS.  This spin up
     could result in less efficient internal mixing and a
     longer-lasting presence of photospheric lithium
     \citep{eggenberger_impact_2012}.
\end{itemize}

The existence of the halo itself is not particularly surprising.  A
star with a 1\kms\ velocity difference from the cluster average will
move away from the center at a rate of 1\,pc\,Myr$^{-1}$.  For a
$\sim$100\,Myr cluster, a halo with characteristic size $\sim$100\,pc
is therefore expected, given that the typical velocity dispersions of
open clusters are of order kilometers per second.

On the other hand, our ability to reliably identify members of these
halos is rather surprising.  These stars have remained hidden in the
background of the Galaxy throughout centuries of modern astronomy.
Their discovery represents an important step toward understanding
the dispersal of open clusters.
Combined with wide-field photometric surveys, the stars themselves
enable advances in studies of stellar
angular momentum evolution.
Combined with wide-field spectroscopic
surveys \citep[{\it e.g.},][]{kollmeier_2017}, the halos could also
improve our
understanding of the chemical evolution of low-mass stars.
Finally, the halo stars provide an expanded sample of
age-dated stars to search for planets \citep[{\it
e.g.},][]{newton_2021}, which may help in benchmarking our understanding of
planetary evolution.


\acknowledgements
\raggedbottom

L.G.B{.} is grateful to G.~Zhou, B{.}~Tofflemire, A{.}~McWilliam,
E{.}~Newton, M{.}~Kounkel, A{.}~Kraus, L{.}~Hillenbrand, and
K{.}~Hawkins for the discussions on young stars, rotation, and lithium
that encouraged this analysis.
L.G.B{.} and J.H{.} acknowledge support by the TESS GI Program, programs
G011103 and G022117, through NASA grants 80NSSC19K0386 and
80NSSC19K1728.
L.G.B{.} was also supported by a Charlotte Elizabeth Procter Fellowship
from Princeton University.
This study was based in part on observations at Cerro Tololo
Inter-American Observatory at NSF's NOIRLab (NOIRLab Prop{.} ID
2020A-0146; 2020B-0029 PI: Bouma), which is managed by the
Association of Universities for Research in Astronomy (AURA) under a
cooperative agreement with the National Science Foundation.
%
%
This paper also includes data collected by the TESS mission, which are
publicly available from the Mikulski Archive for Space Telescopes
(MAST).
Funding for the TESS mission is provided by NASA's Science Mission
directorate.
We thank the TESS Architects (G.~Ricker, R.~Vanderspek, D.~Latham,
S.~Seager, J.~Jenkins) and the many TESS team members for their
efforts to make the mission a continued success.

\software{
  \texttt{astrobase} \citep{bhatti_astrobase_2018},
  \texttt{astropy} \citep{astropy_2018},
  \texttt{astroquery} \citep{astroquery_2018},
  \texttt{cdips-pipeline} \citep{bhatti_cdips-pipeline_2019},
  \texttt{corner} \citep{corner_2016},
	\texttt{gala} \citep{gala,PriceWhelan_2017_gala_zenodo},
  \texttt{IPython} \citep{perez_2007},
  \texttt{matplotlib} \citep{hunter_matplotlib_2007}, 
  \texttt{numpy} \citep{walt_numpy_2011}, 
  \texttt{pandas} \citep{mckinney-proc-scipy-2010},
  \texttt{scikit-learn} \citep{scikit-learn},
  \texttt{scipy} \citep{jones_scipy_2001},
  \texttt{wotan} \citep{hippke_wotan_2019}.
}
\ 

\facilities{
 	{\it Astrometry}:
 	Gaia \citep{gaia_collaboration_gaia_2018,gaia_collaboration_2020_edr3}.
 	{\it Imaging}:
    Second Generation Digitized Sky Survey. 
 	{\it Spectroscopy}:
	CTIO1.5$\,$m~(CHIRON; \citealt{tokovinin_chironfiber_2013}),
	AAT~(HERMES; \citealt{lewis_2002_hermers_2df,sheinis_2015_hermes}),
 	VLT:Kueyen~(FLAMES; \citealt{pasquini_2002}).
 	{\it Photometry}:
 	TESS \citep{ricker_transiting_2015}.
}

\begin{deluxetable*}{lll}
    

\tabletypesize{\small}


\tablecaption{Rotation periods and lithium equivalent widths for
\nkinematic\ candidate NGC\,2516 members.}
\label{tab:maintable}


\tablehead{
  \colhead{Parameter} &
  \colhead{Example Value} &
  \colhead{Description}
}

%
\startdata
               \texttt{source\_id} & 5290666991552265088 &                                                                                Gaia DR2 source identifier. \\
         \texttt{source\_id\_edr3} & 5290666991552265088 &                                                                               Gaia EDR3 source identifier. \\
                 \texttt{in\_SetA} &                True &                                    In Set $\mathcal{A}$ (LSP$>$0.08, P$<$15d, nequal$==$0, nclose$\geq$1). \\
                 \texttt{in\_SetB} &                True & In Set $\mathcal{B}$ (Set $\mathcal{A}$ and below $P_\mathrm{rot}-(G_\mathrm{BP}$-$G_\mathrm{RP})_0$ cut). \\
         \texttt{n\_cdips\_sector} &                   7 &                                                            Number of TESS sectors with CDIPS light curves. \\
                   \texttt{period} &            7.733189 &                                                                           Lomb-Scargle best period [days]. \\
                   \texttt{lspval} &            0.505373 &                                                            Lomb-Scargle periodogram value for best period. \\
                   \texttt{nequal} &                   0 &                                                 Number of stars brighter than the target in TESS aperture. \\
                   \texttt{nclose} &                   1 &                                                   Number of stars with $\Delta T > 1.25$ in TESS aperture. \\
                   \texttt{nfaint} &                   1 &                                                    Number of stars with $\Delta T > 2.5$ in TESS aperture. \\
                       \texttt{ra} &          119.494256 &                                                                            Gaia DR2 right ascension [deg]. \\
                      \texttt{dec} &          -61.060819 &                                                                                Gaia DR2 declination [deg]. \\
               \texttt{ref\_epoch} &              2015.5 &                                                       Reference epoch for right ascension and declination. \\
                 \texttt{parallax} &            2.462602 &                                                                                   Gaia DR2 parallax [mas]. \\
          \texttt{parallax\_error} &            0.017613 &                                                                       Gaia DR2 parallax uncertainty [mas]. \\
                     \texttt{pmra} &           -5.366865 &                                        Gaia DR2 proper motion $\mu_\alpha \cos \delta$ [mas$\,$yr$^{-1}$]. \\
                    \texttt{pmdec} &           10.959816 &                                                    Gaia DR2 proper motion $\mu_\delta$ [mas$\,$yr$^{-1}$]. \\
       \texttt{phot\_g\_mean\_mag} &           14.592845 &                                                                                    Gaia DR2 $G$ magnitude. \\
      \texttt{phot\_bp\_mean\_mag} &           15.189163 &                                                                        Gaia DR2 $G_\mathrm{BP}$ magnitude. \\
      \texttt{phot\_rp\_mean\_mag} &           13.867151 &                                                                        Gaia DR2 $G_\mathrm{RP}$ magnitude. \\
         \texttt{radial\_velocity} &                 NaN &                                                    Gaia DR2 heliocentric radial velocity [km$\,$s$^{-1}$]. \\
  \texttt{radial\_velocity\_error} &                 NaN &                                                     Gaia DR2 radial velocity uncertainty [km$\,$s$^{-1}$]. \\
               \texttt{subcluster} &                core &                                                                 Is star in core (CG18) or halo (KC19+M21)? \\
                 \texttt{in\_CG18} &                True &                                                                    Star in \citet{cantatgaudin_gaia_2018}. \\
                 \texttt{in\_KC19} &                True &                                                                   Star in \citet{kounkel_untangling_2019}. \\
                  \texttt{in\_M21} &                True &                                                                             Star in \citet{meingast_2021}. \\
               \texttt{(Bp-Rp)\_0} &            1.187693 &              Gaia $G_\mathrm{BP}$-$G_\mathrm{RP}$ color, minus $E$($G_\mathrm{BP}$-$G_\mathrm{RP}$)=0.1343 \\
            \texttt{is\_phot\_bin} &               False &                                                                True if $>0.3$ mag above cluster isochrone. \\
           \texttt{is\_astrm\_bin} &               False &                                                                              True if Gaia EDR3 RUWE > 1.2. \\
                     \texttt{ruwe} &            0.954835 &                                                                                            Gaia EDR3 RUWE. \\
      \texttt{Li\_EW\_mA\_GaiaESO} &             144.259 &                                       Gaia-ESO Li doublet equivalent width, including the Fe blend [m\AA]. \\
\texttt{Li\_EW\_mA\_perr\_GaiaESO} &               6.301 &                                                           Gaia-ESO Li doublet EW upper uncertainty [m\AA]. \\
\texttt{Li\_EW\_mA\_merr\_GaiaESO} &                 5.0 &                                                           Gaia-ESO Li doublet EW lower uncertainty [m\AA]. \\
        \texttt{Li\_EW\_mA\_GALAH} &                 NaN &                                          GALAH Li doublet equivalent width, including the Fe blend [m\AA]. \\
  \texttt{Li\_EW\_mA\_perr\_GALAH} &                 NaN &                                                              GALAH Li doublet EW upper uncertainty [m\AA]. \\
  \texttt{Li\_EW\_mA\_merr\_GALAH} &                 NaN &                                                              GALAH Li doublet EW lower uncertainty [m\AA]. \\
\enddata


\tablecomments{
Table~\ref{tab:maintable} is published in its entirety in a
machine-readable format.  One entry is shown for guidance regarding
form and content.  This table is a concatenation of all candidate
NGC\,2516 members reported by \citetalias{cantatgaudin_gaia_2018},
\citetalias{kounkel_untangling_2019}, and \citetalias{meingast_2021}
based on the Gaia DR2 data.  Different levels of purity and
completeness can be achieved using different cuts on photometric
periods, periodogram powers, and lithium eqiuvalent widths.  Sets
$\mathcal{A}$ and $\mathcal{B}$ provide two possible levels of
cleaning (see Section~\ref{subsubsec:cluster}).  When the
target star is the only star present in the TESS aperture,
\texttt{nequal}$=0$, \texttt{nclose}$=1$, and \texttt{nfaint}$=1$.
The light curves are available at
\url{https://archive.stsci.edu/hlsp/cdips}.
Supplementary plots enabling analyses of individual stars
are available at \url{https://lgbouma.com/notes}.
}
\vspace{-0.5cm}
\end{deluxetable*}

\clearpage
\bibliographystyle{yahapj}                            
\bibliography{bibliography} 

\appendix
\section{Clustering Methods and Outcomes}
\label{app:clustering}

\begin{figure}[b]
	\begin{center}
		\leavevmode
		\includegraphics[width=0.3\textwidth]{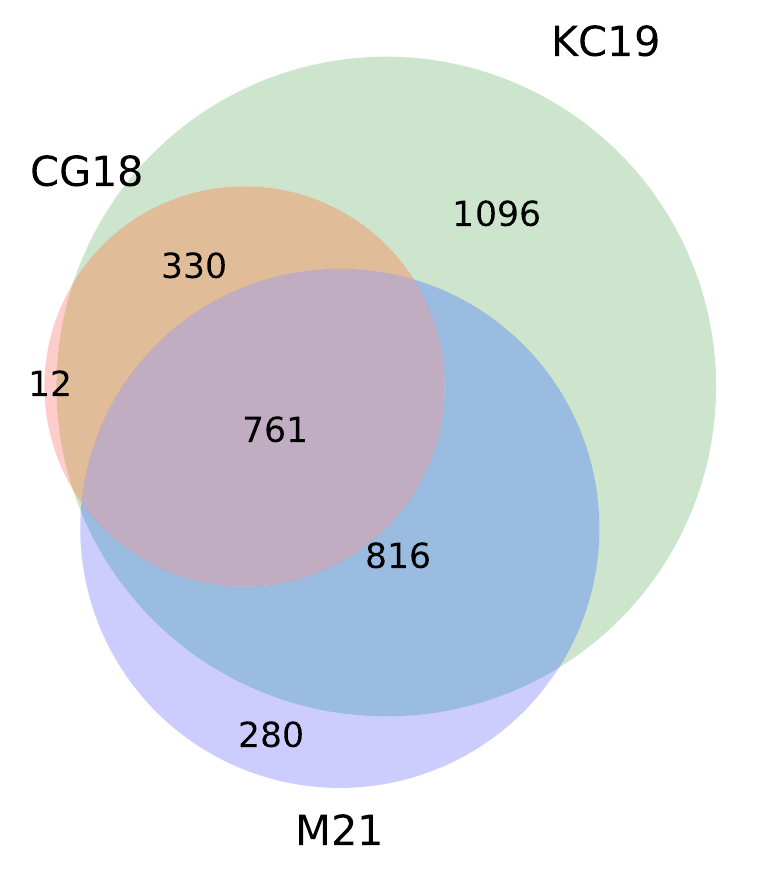}
	\end{center}
	\vspace{-0.7cm}
  \caption{ {\bf Different clustering techniques yield different
  candidate members of \cn.} \nkinematic\ unique candidate cluster
  members found using three different techniques are considered in
  this study.  \citetalias{cantatgaudin_gaia_2018}:
  \citet{cantatgaudin_gaia_2018},
  \citetalias{kounkel_untangling_2019}:
  \citet{kounkel_untangling_2019}, \citetalias{meingast_2021}:
  \citet{meingast_2021}.
  \label{fig:venn}
	}
\end{figure}

Figure~\ref{fig:venn} is a Venn diagram of
the three membership catalogs concatenated in this study.  99\% of the
\citetalias{cantatgaudin_gaia_2018} sample overlaps with
\citetalias{kounkel_untangling_2019}, \citetalias{meingast_2021}, or
both.  By comparison, 36\% of the \citetalias{kounkel_untangling_2019}
sample, and 15\% of the \citetalias{meingast_2021} sample do not
overlap with any of the other catalogs.  The data---Gaia DR2---used by
all the studies was the same.  What are the differences in methodology
that produce these different membership lists?

\citetalias{cantatgaudin_gaia_2018} applied a procedure that yielded
what we colloquially call ``the core''.  Their membership assignment
algorithm was to first query a Gaia DR2 cone around the previously
reported $\{\alpha,\delta\}$ of the cluster center, and within $\pm
0.5$\,mas of its previously reported parallax.  The outer radius of
their cone was the previously reported angular radius of the cluster
(0.71$^\circ$; \citealt{Kharchenko_et_al_2013}).  No proper motion cut
was applied.  \citetalias{cantatgaudin_gaia_2018} then applied an
unsupervised classification scheme to $G<18$\,mag stars within this
cone (UPMASK; \citealt{kronemartins_upmask_2014}).  The UPMASK
algorithm first performs a $k$-means clustering on the astrometric
parameters, $\{\mu_{\alpha'}, \mu_\delta, \pi\}$.  A ``veto''
step is then applied to assess whether the groups of stars output from the
$k$-means clustering are more concentrated than a uniform
distribution, by comparing the total branch lengths of their
respective minimum spanning trees.  This binary flag is converted to a
membership probability by redrawing values for $\{\mu_{\alpha'},
\mu_\delta, \pi\}$ using the reported uncertainties and
covariances.  The final membership probability for any given star is
then the fraction of draws for which the star passes the veto step.
In the case of \cn, the resulting radius within which half of the
cluster members were found was 0.50$^\circ$.  We included all
\citetalias{cantatgaudin_gaia_2018} \cn\ members with reported
membership probability exceeding 10\% in our list of candidate cluster
members.

\citetalias{kounkel_untangling_2019} applied a different unsupervised
clustering method using the 5D Gaia DR2 positions and proper motions.
\citetalias{kounkel_untangling_2019} began with a sample of
$\approx$$2\times 10^7$ stars, mostly within $\approx 1$\,kpc and with
$G\lesssim18$\,mag.  They then applied the hierarchical density-based
spatial clustering of applications with noise algorithm (HDBSCAN;
\citealt{campello_hierarchical_2015}, \citealt{mcinnes_hdbscan_2017})
on the entire sample,  setting the minimum number of stars per cluster
to be 40.  \citetalias{kounkel_untangling_2019} then iterated over a
few different cutoff parallaxes to improve sensitivity to structures
with distances from the Sun ranging between $\approx$100\,pc and
$\approx$1000\,pc.  They then fitted ages to the resulting clusters
using two different methods: isochronal, and a convolutional neural
network.  Regarding membership proabilities,
\citetalias{kounkel_untangling_2019} reported the binary ``member'' or
``not'' from HDBSCAN.
We included all \citetalias{kounkel_untangling_2019} \cn\ members in
our list of candidate cluster members.

Finally, \citetalias{meingast_2021} considered ten known open clusters
within the nearest kiloparsec. One of these clusters was \cn.  Their
initial  list of $\approx$$3\times10^7$ stars was similar to that of
\citetalias{kounkel_untangling_2019}, mainly requiring well-behaved
astrometry and photometry from Gaia.  Given the mean positions and
velocities of the clusters determined by
\citetalias{cantatgaudin_gaia_2018}, \citetalias{meingast_2021} then
selected all stars with proper-motion correct tangential velocities
within 1.5\,\kms\ of the cluster mean.  They then applied DBSCAN
\citep{ester_density-based_1996} on both the observed galactic
$\{X,Y,Z\}$ coordinates, as well as to a separate deconvolved spatial
coordinate distribution discussed in their Section~3.3.  The results
of their procedure, as well as those of
\citetalias{cantatgaudin_gaia_2018} and
\citetalias{kounkel_untangling_2019}, are visible at the website of
S.~Meingast:
\url{https://homepage.univie.ac.at/stefan.meingast/coronae.html}.
Their reported membership proabilities were binary, similar to those
of \citetalias{kounkel_untangling_2019}.  All of the
\citetalias{meingast_2021} \cn\ members were included in our list of
candidate cluster members.

\section{Considerations for Removing Systematic Variability from TESS Light Curves}
\label{app:detrending}

\begin{figure}[btp]
	\begin{center}
		\leavevmode
		\includegraphics[width=\textwidth]{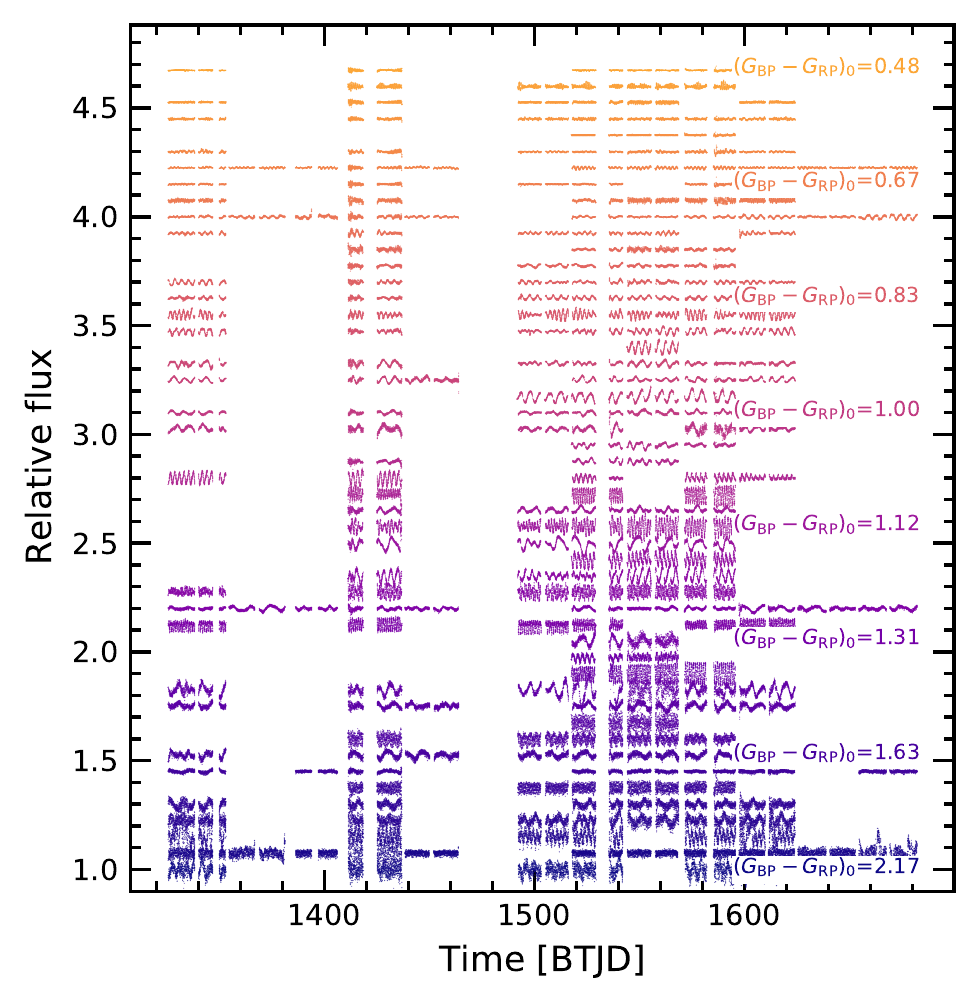}
	\end{center}
	\vspace{-1cm}
  \caption{ {\bf \cn\ light curves sorted in order
  of stellar color.} Fifty stars are randomly selected from Set
  $\mathcal{B}$ (see Section~\ref{subsubsec:cluster}). The light
  curves were detrended and cleaned as discussed in
  Appendix~\ref{app:detrending}.  Each light curve is separated by
  7.5\% in relative flux.
  \label{fig:lightcurves}
	}
\end{figure}

In detrending before searching for stellar rotation signals, our goal
was to preserve astrophysical variability, while removing systematic
variability.  One particular concern for the TESS light curves is
systematic variability at the timescale of the 14-day satellite orbit,
mostly induced by scattered light from the Earth and Moon.

We opted to use a variant of the principal component analysis (PCA)
approach discussed by \citet{bouma_cdipsI_2019}. This PCA approach
uses a set of ``trend stars'' selected from across each CCD according
using ad-hoc heuristics that on average lead the trend star light
curves to be dominated by systematic variability.  The resulting
principal component vectors, also referred to as the eigenvectors, are
rank-ordered by the degree of variance that they predict in the
training set of trend stars.

We then posit that any given target star's light curve is described as
a linear combination of the eigenvectors.  We also considered the
inclusion of additional systematic vectors that could affect the light
curve, discussed below.  These can be treated as additional
``features'' in the linear model.  To determine the coefficients of
the linear model after the full set of eigenvectors (plus optional
``sytematic'' vectors) had been asssembled,  we explored two possible
methods: ordinary least squares, and ridge regression. Ridge
regression is the same as ordinary least squares, except it includes
an $\ell^2$ norm with a regularization coefficient. The regularization
coefficient that best applied for any given target light curve was
determined using a cross-validation grid search, through
\texttt{sklearn.linear\_modelRidgeCV} \citep{scikit-learn}.  Each
target light curve was mean-subtracted and normalized by its standard
deviation, as were the eigenvectors. The linear problem was then
solved numerically, and the light curve was reconstructed by re-adding
the original mean, and re-multiplying by the standard deviation to
ensure that the variance of the light curve did not change.

We found that the choice of using ordinary least squares versus ridge
regression did not significantly affect the resulting light
curves. In other words, the inclusion (or lack thereof) of a
regularization term did not strongly alter the best-fitting
coefficients.  In the spirit of keeping it simple, we opted for
ordinary least squares.  A few other choices seemed to be more
important:

\begin{itemize}
  \item {\it To smooth, or to not smooth the eigenvectors}.
    If the systematic trends are smooth in time, the eigenvectors should be as well. They should not
    contain residuals from {\it e.g.}, eclipsing binaries 
    in the template set, and they should also not be
    intrinsically noisier than the target star. If either of these is
    the case, extra
    variability can be introduced into the ``detrended'' light curves.  To address
    this problem, we opted to smooth the eigenvectors using a
    sliding time-windowed filter (with a "biweight" weight scheme, implemented
    in \texttt{wotan} by \citealt{hippke_wotan_2019}).
    One issue with this is that systematic sharp features (``spikes'')
    are no longer captured, and could be present in the detrended light
    curves. Since they can be filtered out in postprocessing
    ({\it e.g.}, using rolling outlier rejection), we prefer this
    approach to having systematic features being injected by the
    PCA detrending.
  \item {\it How many eigenvectors to use}.
    A larger number always leads to greater whitening.  In
    \citet{bouma_cdipsI_2019}, we performed a factor analysis
    cross-validation to determine the number of eigenvectors to use.
    The typical number adopted based on this analysis was 10 to 15.
    While this approach should in theory prevent over-fitting, in our
    experience, for stellar rotation it still often lead to distorts
    the signals, especially for rotation signals with small amplitudes
    and periodicities of $\gtrsim 3$ days.  Shorter signals typically
    are not distorted, since the eigenvectors do not contain the
    high-frequency content that leads to the distortions.  For the
    present analysis, we therefore set the maximum number of
    eigenvectors to be 5.
  \item {\it Which supplementary systematics vectors to use}.  We
    considered using the \texttt{BGV}, \texttt{CCDTEMP}, \texttt{XIC},
    \texttt{YIC}, and \texttt{BGV} vectors, packaged with the CDIPS
    light curves. We found that the background value measured in an
    annulus centered on the aperture, \texttt{BGV}, tended to produce
    the best independent information from the PCA eigenvectors, and so
    we adopted it as our only ``supplementary'' trend vector.  We
    opted to not smooth it, assuming that it would provide direct
    complement to the smoothed PCA vectors.
\end{itemize}

After all these considerations, for every target star, we ultimately
decorrelated the raw (image-subtracted and background-subtracted)
light curve using a linear model with ordinary least squares.  The
vectors used in the model were the 1 unsmoothed background flux
vector, and 5 smoothed PCA eigenvectors.  Figure~\ref{fig:lightcurves}
shows fifty light curves randomly drawn from the resulting Set
$\mathcal{B}$.  A supplementary figure set that enables inspection of
each individual star is available at \url{https://lgbouma.com/notes}.

\section{Rotation Periods for Comparison Field Stars}
\label{app:compstar}

\begin{figure}[t]
	\begin{center}
		\leavevmode
    \includegraphics[width=0.69\textwidth]{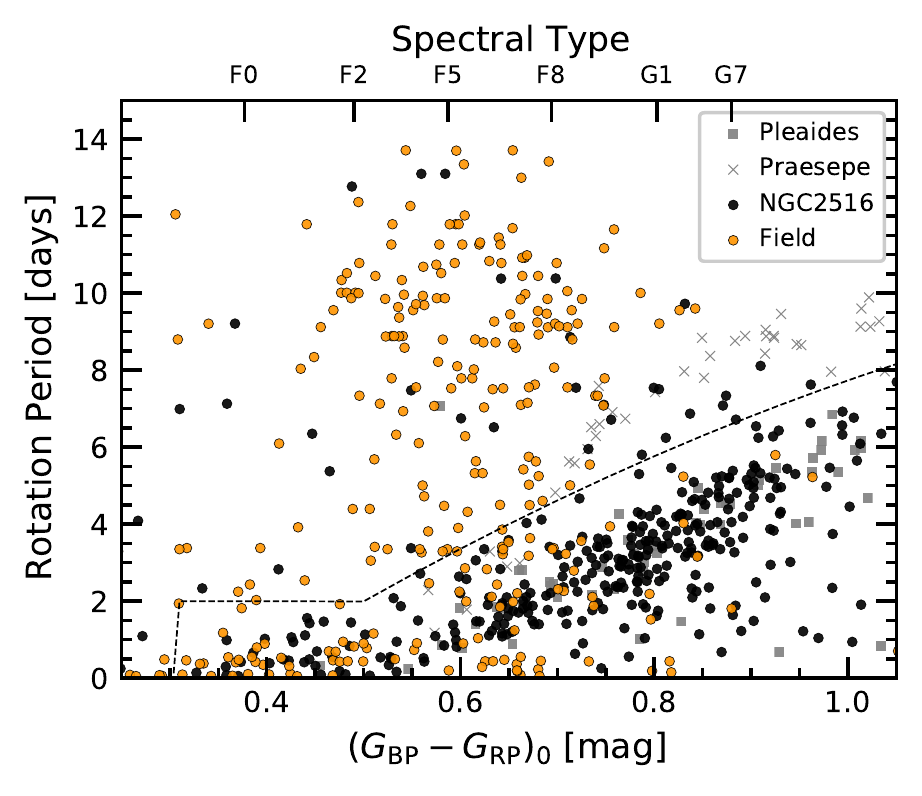}
	\end{center}
	\vspace{-0.7cm}
	\caption{ {\bf Rotation in \cn\ compared to the field.}
  Stars with Gaia $G_\mathrm{RP}<13$\,mag in the cluster and field samples are
  displayed; fainter field stars were not analyzed.  Field
  stars (orange) show a different rotation period distribution than
  the kinematically selected \cn\ members (black).  The field and \cn\
  stars are both subject to the same 
  period-measurement and cleaning procedures.
  The dashed line is as in Figure~\ref{fig:rot}.
  \label{fig:compstar}
	}
\end{figure}

To verify the significance of the trends seen in Figures~\ref{fig:rot}
and~\ref{fig:binarity}, we searched the CDIPS calibration light curves
for rotation periods.  This light curve database comprises all $G_\mathrm{RP}<13$
stars that fell on TESS silicon. The need for separate field and
cluster star magnitude limits is based on our processing capabilities and is discussed
by \citet{bouma_cdipsI_2019}.  Over the southern sky (Sectors 1-13 of
TESS), this corresponded to a sample of \ncalibration\ stars.
Cross-matching these against the \nnbhd\ randomly drawn stars in the
neighborhood of \cn\ yielded \nnbhdcalibstar\ unique stars.  The
magnitude cut of $G_\mathrm{RP}<13$ at the distances of the neighborhood sample
corresponds to an extinction-corrected color cutoff of $\bpmrpo
\lesssim 0.80$, or spectral types $\lesssim$G1V.  This reaches
sufficiently far down the slow sequence to enable a comparison against
the cluster star sample.

We performed the same light curve stitching and period-search
procedure discussed in the text on the field stars.
Within $0.4<\bpmrpo<0.8$, the same requirements
for crowding resulted in \ncompstardenominator\ field stars
for which rotation periods could have been detected.  Imposing the
same Lomb-Scargle power cutoff (${\rm LSP}>0.08$) and color-period cutoff as that used to
define Set $\mathcal{B}$ yielded \ncompstarnumerator\ period
detections (\ncompfrac).  Within the same color range, \nautovscompstarnumerator\ of
\nautovscompstardenominator\ kinematically
identified candidate cluster members yielded period detections
(\nautofrac).  The period--color distributions
are also quite different: the candidate
cluster members show an overdensity coincident with the expected
gyrochronal age, while the field stars show a much broader rotation
period distribution (Figure~\ref{fig:compstar}).  This confirms our
expectation that our period measurement procedure is not biased in any
way that could produce the features we observe in \cn.

\section{Projection Effects for Tangential Velocities}
\label{app:vproj}

\begin{figure}[t]
	\begin{center}
		\leavevmode
		\subfloat{
			\includegraphics[width=0.49\textwidth]{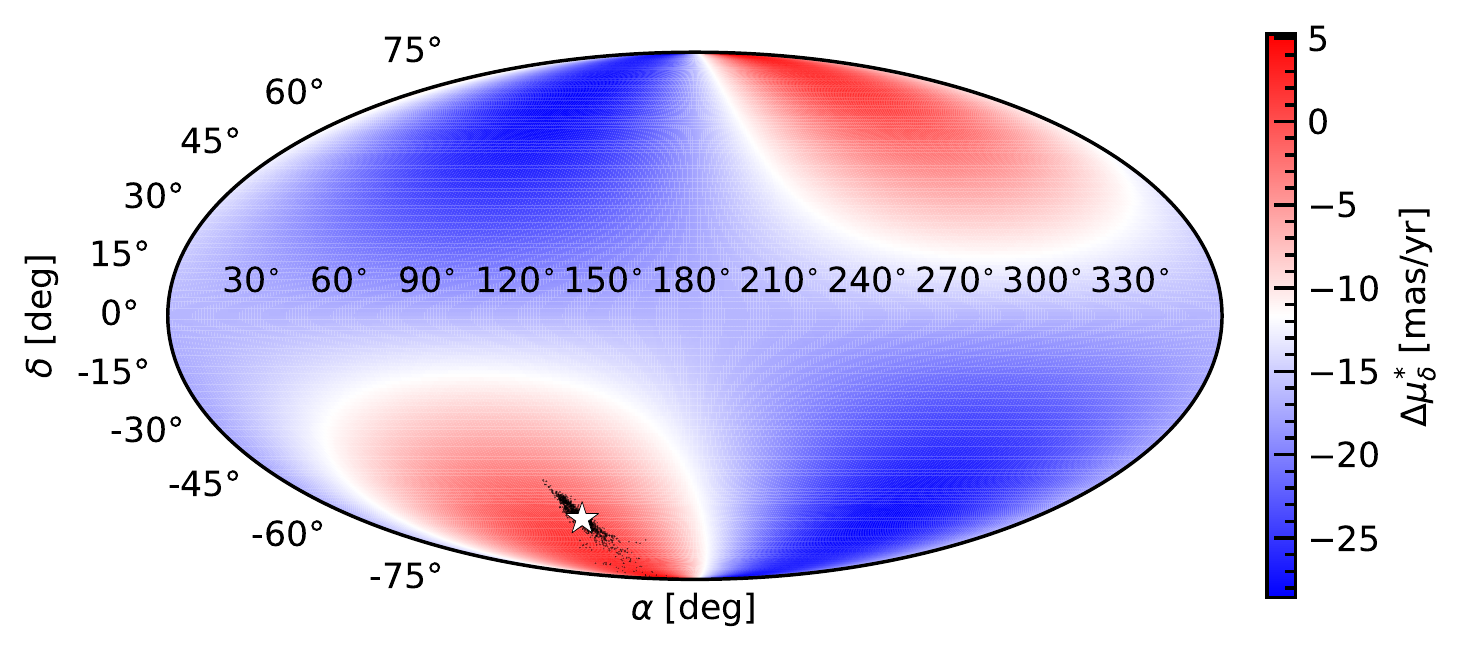}
			\includegraphics[width=0.49\textwidth]{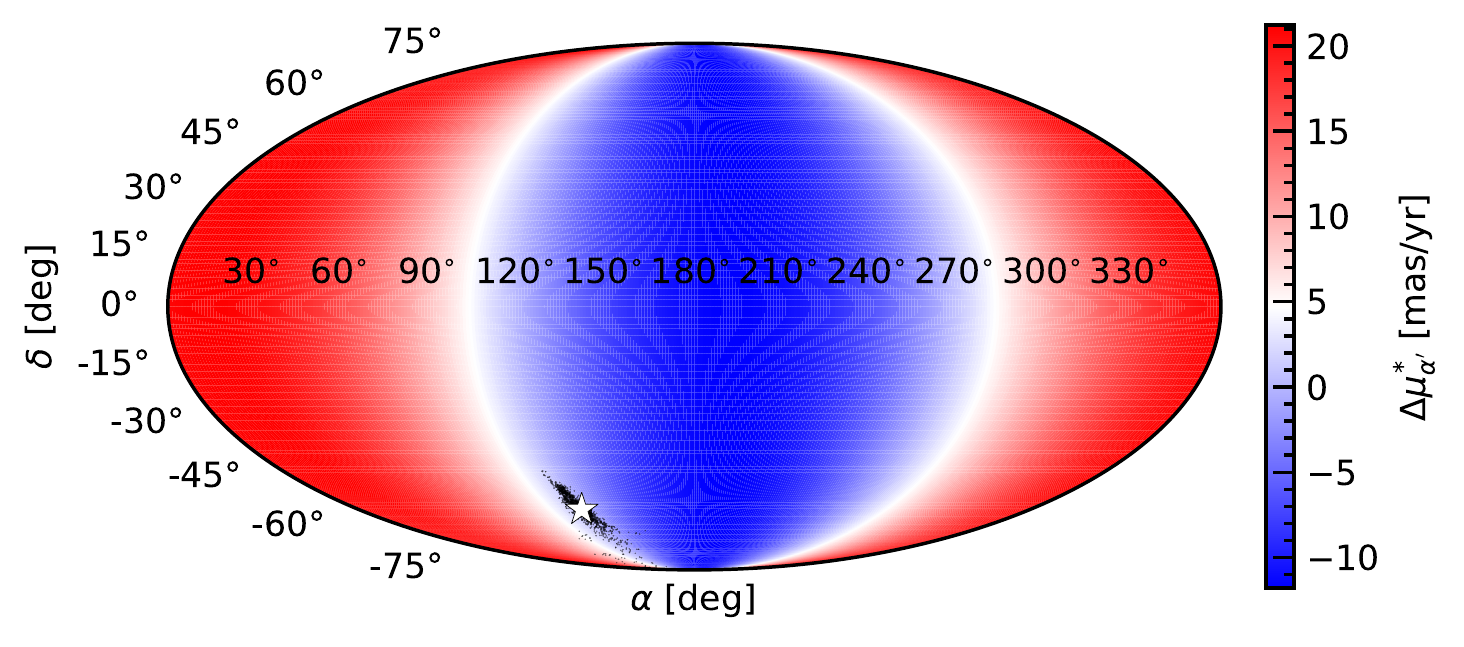}
		}
	\end{center}
	\vspace{-0.5cm}
  \caption{ {\bf Projection effects for co-moving populations across
  the sky.} The maps are colored by the proper motion difference a star
  co-moving with \cn\ would exhibit across the equatorial sphere.
  Actual positions of candidate \cn\ members are shown with points;
  the star denotes the cluster center.
  The {\it left} and {\it right} panels show the proper motion correction in
  the declination and right ascension directions.
	\label{fig:vproj}
	}
\end{figure}

To compare any given star's proper motion against the cluster's mean,
one must consider the position of the star on the sky.  This is
because two stars sharing identical Cartesian velocities will have
different velocities when projected on the sky, as was emphasized by
\citet{meingast_2021}.
Figure~\ref{fig:vproj} shows the magnitude of the correction in
equatorial tangential velocities.

%
%

\section{Positions and Kinematics of the Rotators}
\label{app:gaia6d_x_rotn}

\begin{figure*}[tp]
	\begin{center}
		\leavevmode
		\includegraphics[width=0.95\textwidth]{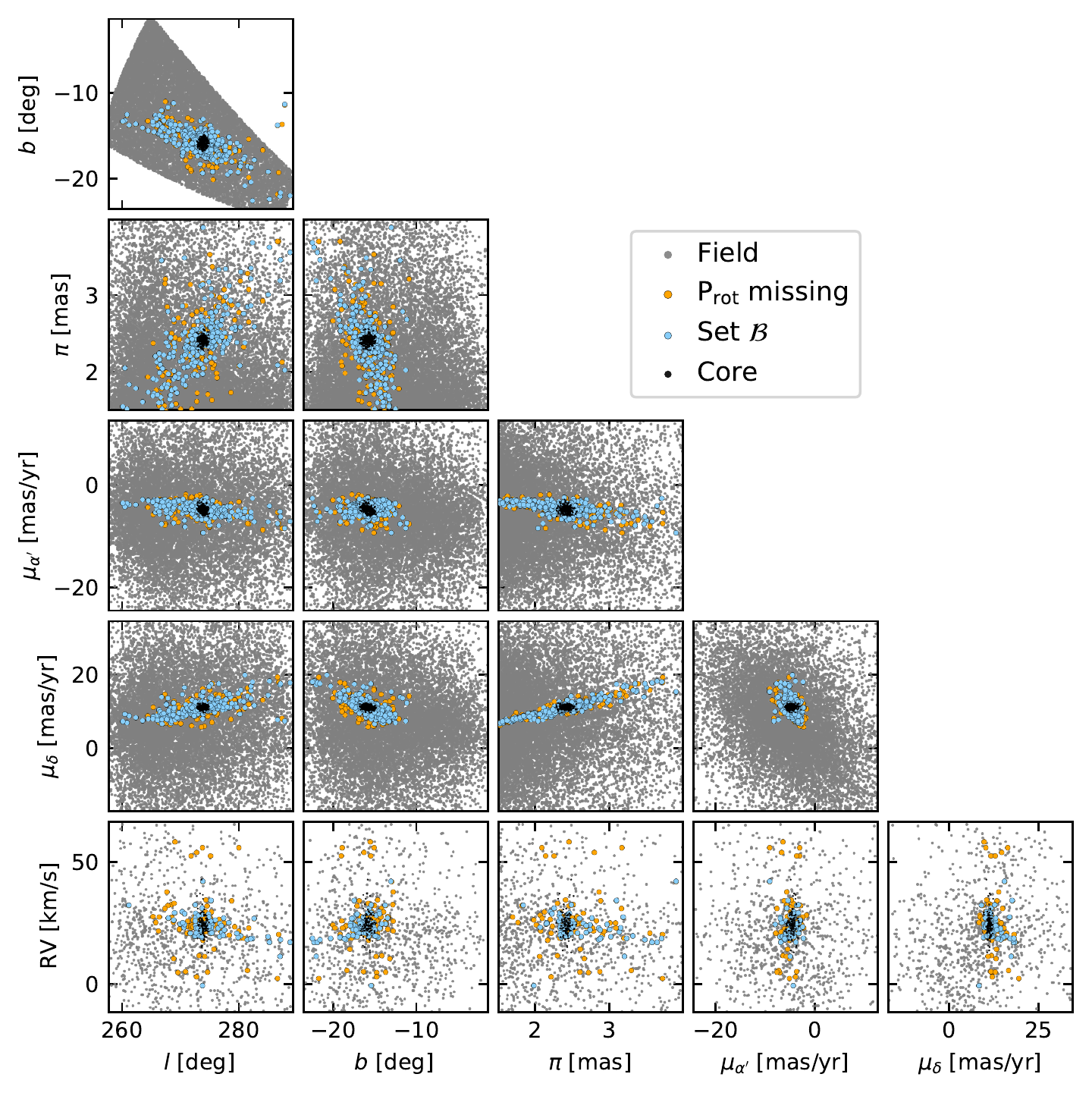}
	\end{center}
	\vspace{-0.7cm}
  \caption{ {\bf Core and halo of NGC\,2516 in position and
  velocity space, cross-matched against the rotators.} The plotted
  cluster members are those with $0.5<\bpmrpo<1.2$ that meet the
  crowding restrictions described in Section~\ref{subsec:tess}: they
  should have been sufficiently bright and non-crowded to enable
  rotation period detection. Stars in Set $\mathcal{B}$ are shown in blue; those which should have
  had rotation periods detected but did not are in orange.
  We caution that the appearance of fewer
  non-rotators being present toward the core is due the layering of
  the plot: quantitatively, stars toward the cluster center do not all
  show rotation periods in our analysis (see
  Figure~\ref{fig:physical_x_rotn}).
  \label{fig:gaia6d_x_rotn}
	}
\end{figure*}

Figure~\ref{fig:gaia6d_x_rotn} is an alternative visualization
of the data shown in Figure~\ref{fig:physical_x_rotn}.  The rotation
periods in this diagram correspond to ``Set $\mathcal{B}$'' as
described in Section~\ref{subsec:tess}.  One issue with the
visualization as displayed however is that the layering hides the 
non-rotators toward the cluster center: we did not detect rotation
periods for roughly one in four stars at the cluster center (see
upper-right panel in Figure~\ref{fig:physical_x_rotn}).  
%
%
%
%
%
%
%
%
%

\listofchanges

\end{document}